\RequirePackage{fix-cm}
\documentclass[twocolumn]{svjour3}       
\smartqed  
\usepackage[pdftex]{graphicx}
\usepackage{color}
\usepackage{amsmath}
\usepackage{amsfonts}
\begin{document} 

\title{From materials to systems: a multiscale analysis of nanomagnetic switching
}


\author{Yunkun~Xie \and Jianhua~Ma \and Samiran~Ganguly \and Avik~W.~Ghosh 
}


\institute{ Department of Electrical and Computer Engineering, University of Virginia, Charlottesville,
VA, 22903 \\
              \email{yx3ga@virginia.edu}           
}

\date{Received: date / Accepted: date}

\maketitle

\begin{abstract}
With the increasing demand for low-power electronics, nanomagnetic devices have emerged as strong potential candidates to complement present day transistor technology. A variety of novel switching effects such as spin torque and giant spin Hall offer scalable ways to manipulate nano-sized magnets. However, the low intrinsic energy cost of switching spins is often compromised by the energy consumed in the overhead circuitry in creating the necessary switching fields. Scaling brings in added concerns such as the ability to distinguish states (readability) and to write information without spontaneous backflips (reliability). A viable device must ultimately navigate a complex multi-dimensional material and design space defined by volume, energy budget, speed and a target read-write-retention error. 
 In this paper, we review the major challenges facing nanomagnetic devices and present a multi-scale computational framework to explore possible innovations at different levels (material, device, or circuit), along with a holistic understanding of their overall energy-delay-reliability tradeoff. 
\keywords{nanomagnetics, STT-MRAM, computational spintronics, spin logic, neuromorphic spintronics.}
\end{abstract}

\section{Introduction}
Power dissipation is a major concern in fast growing mobile applications where computations must be performed with a limited power budget. One of the fundamental issues with electronic devices is their need to move several electrons to charge/discharge a gate or a transmission line. The power dissipation in this process has been shown to be about $Nk_BT\ln(1/p)$ where $T$ is the temperature, $k_B$ is Boltzmann's constant, $N$ is the number of electrons/holes transferred and $p$ is the switching error probability of the device \cite{ghosh2015nanoelectronics}. Nano-magnetic devices, on the other hand, seem to consume much less switching energy due to the `self-correcting' effect \cite{salahuddin2007interacting}, where their internal exchange coupling bundles the electron spins and allows them to act as a single entity while simultaneously providing resilience against noise. The minimum energy consumption from the switching of a nanomagnet (i.e., a collection of spins) is estimated to be on the order of $\sim k_BT\ln(1/p)$ which is significantly lower than that in electronic devices for comparable number of information carriers \cite{salahuddin2007interacting}. 

In the past decade, nanomagnetic devices have started to carve out a space for themselves as good candidates for memory applications due to the nonvolatility of magnets. The spin transfer torque magnetic random access memory (STT-MRAM) has been touted as the solid-state `universal memory' device \cite{huai_spin-transfer_2008} that can flatten the register-cache-RAM-storage hierarchy due to its high scalability, almost limitless endurance, back end of line (BEOL) integrability in standard complementary metal oxide semiconductor (CMOS) manufacturing process, and relatively high speed and low dissipation. Uncertain future of flash memory \cite{grupp2012bleak} has opened a possible market for STT-MRAM to be the choice of storage in embedded and mobile applications, and perhaps even in cloud computing as large scale, high speed, low power data-caches. Though challenges remain, especially in scaling up the production which ultimately lowers the barrier of entry into the market for STT-MRAMs, recent developments \cite{everspin1gb,gf22nm,everspingoflo,samsungvideo} have shown a great promise for the large scale commercialization of the technology.

A rich variety of magnetic switching phenomena have appeared recently on the horizon, such as spin transfer torque, spin-Hall effect, skyrmions, multiferroics, and voltage controlled magnetic anisotropy (VCMA). This has led to the exploration of nanomagnetic devices beyond memory applications. There are efforts to build and evaluate spin based logic devices exploiting unique features such as in-situ memory through non-volatility, stochastic computing, oscillators and magnetic neurons. While it is not clear how many of these ideas will see eventual adoption as functional devices in commercial production, it is imperative to understand the fundamental limits to the performance of these devices, and develop methods to overcoming these challenges based on the physics and properties of the materials and phenomena underlying the nano-magnets and their interactions. 

This paper presents a detailed exposition of an integrated multi-scale approach, starting from materials and going all the way up to circuits and systems that can answer these questions and provide an ideal platform for exploration of this technology.

The rest of the paper is organized as follows. Section 2 lays out a bird's eye view of the fundamental challenges underlying all nano-magnetic devices, and indeed any novel technology in terms of the Read, Write, Reliability metrics of performance. Section 3 introduces the integrated multi-scale DFT-to-SPICE approach. Section 4 presents a Density Functional Theory (DFT) based approach to search for materials with desired properties that improve the Read performance. Section 5 and 6 cover the physics of charge and spin-transport in these devices using the Non-Equlibrium Green's Functions (NEGF) formalism and the interaction of the nano-magnets with charge and spin currents through the stochastic Landau-Lifshitz-Glibert-Slonczewski (LLGS) equation. Section 7 reviews the physics, benefits, and challenges associated with a select few emerging mechanisms for magnetic switching which may complement or supplant the classic spin-torque based switching. Section 8 reviews and connects the various pieces developed in sections 3 through 7 and presents an integrated workflow. Section 9 connects the fundamental physics based approaches developed in the previous sections to SPICE based simulations and illustrates the capabilities of the approach through a few example simulations. Finally we summarize in section 10.

\section{Challenges in nanomagnetic applications}
Over the last decade, nanomagnetic devices, especially STT-MRAMs have seen considerable  progress in optimizing size, switching time, energy efficiency, and overall reliability. Major challenges still exist, such as in the scalability of magnetic tunnel junctions (MTJs). In the meantime, new device ideas have emerged which could potentially reduce energy dissipation nearer to or even below existing charge based technologies, with the added advantage of non-volatility. Understanding the issues for each device in an overall computational architecture requires a holistic analysis, spanning a complex multi-dimensional phase space defined by size, energy budget, desired speed and a target read-write-retention error rate. 
In the following sections, we will review those challenges from the read, write, and reliability points of view.

\subsection{The Read perspective - detecting without disrupting}
\begin{figure}[t]
\centering
\includegraphics[width=8cm]{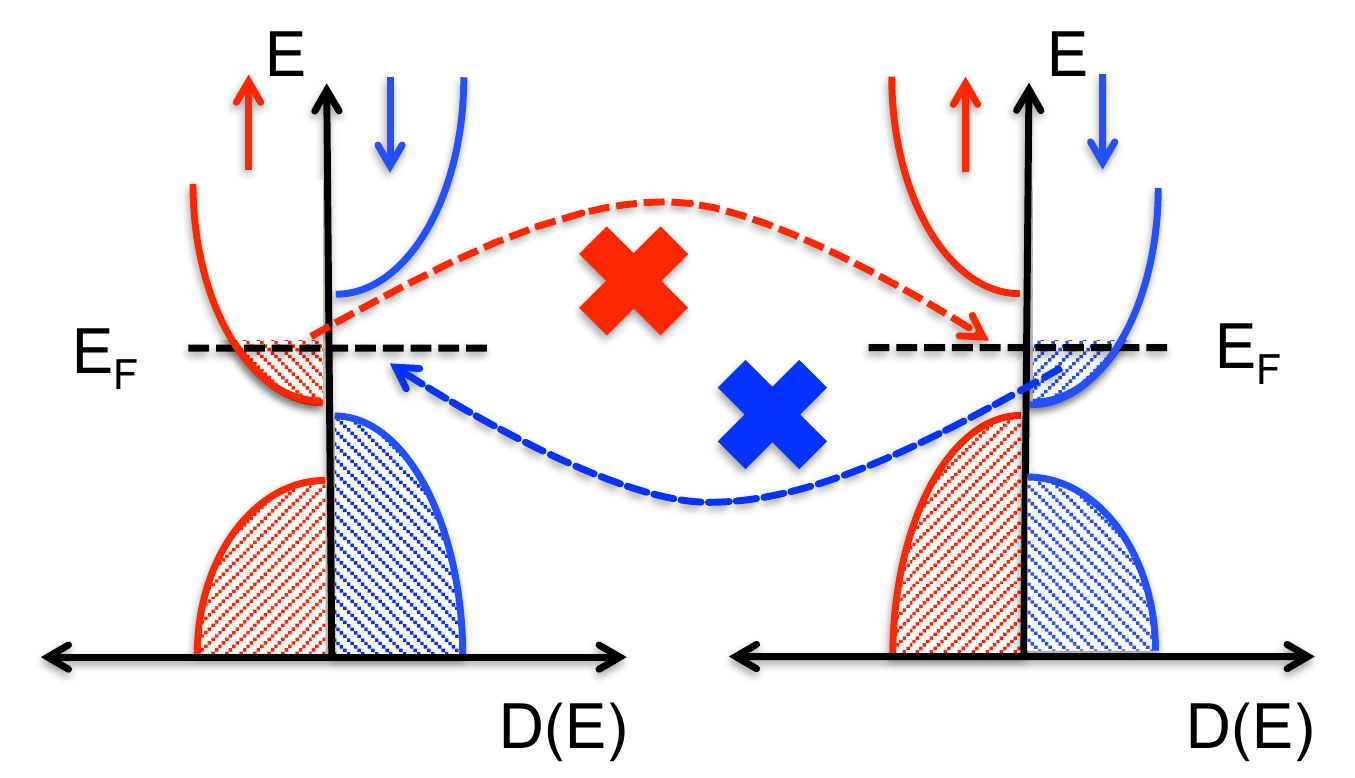}
\caption{Electron tunneling in a MTJ using half-metallic electrodes in the anti-parallel magnetization configuration.}
\label{fig:MTJhalfmetal}
\end{figure}

The ability to read the state of a magnet, say an MTJ, depends on its overall tunnel magnetoresistance (TMR) that quantifies the electrical distinguishability of its two binary states. This ratio would have been infinite if each magnet contained only one spin species (up or down) within the bias range of interest (Fig.~\ref{fig:MTJhalfmetal}). A TMR ratio defined as $\mathrm{TMR}=(R_{ap}-R_p)/R_{p}$ quantifies the readability of the MTJ, where $R_p$ and $R_{ap}$ are the junction resistances in the low resistance (parallel) vs high resistance (antiparallel) configurations. By passing a small current $I_\mathrm{read}$ through the MTJ and comparing the voltage drop against a reference value, we can identify the magnetic configuration of the MTJ. A high TMR ratio is needed to provide enough sensing margin $\Delta V=I_\mathrm{read}R_p\cdot\mathrm{TMR}$, without the need to inject a large  dissipative and disruptive read current that can alter the state of the magnets. Past research has shown high TMR ratios (as high as $\sim 600\%$) for epitaxially grown Fe/MgO\cite{yuasa2004giant} or sputtered FeCo/MgO junctions at room temperature \cite{parkin2004giant,ikeda2008tunnel}, building on the predicted symmetry filtering of MgO and half-metallicity of Fe and FeCo (predictions exceed 4000$\%$ at low bias). However, in state-of-art MTJs where the write current is optimized, the TMR ratio hardly exceeds $100\%$. This degradation may arise from the shrinking thickness of MgO and from spin depolarization by defects like oxygen vacancies sitting near the interfaces between the spacer and the magnetic layers \cite{amiri2011low,mather2006disorder}. Improved fabrication processes plus a comprehensive materials study are needed to enhance the TMR ratio for better readability in MTJs.

\subsection{The Write perspective - spins are cheap but fields are costly}
Although the intrinsic energy dissipation of magnetization switching is low, the energy required to generate the means to flip a magnet is not. For instance, the traditional way of manipulating a magnet is with a magnetic field, which is generated using a current carrying wire which dissipates considerable energy. Furthermore, the field-switching scheme is not scalable; scaling down the magnet reduces its overall magnetic moment, which needs to be compensated by increasing the anisotropy to retain spins for a long enough time. Toggle magnetic switching depends on the anisotropy, needing therefore a larger current for smaller magnets. In comparison, STT switching offers a scalable solution since the critical switching current needed to switch a perpendicular magnet depends on the overall energy barrier (i.e., magnetization $\times$ volume $\times$ anisotropy field), which is held fixed during scaling to achieve a target retention time. As we argue later, the typical current density required to switch a magnet in nanoseconds is on the order of $\sim 1-10\mathrm{MA/cm^2}$, because the spins needed to flip the magnetization need to come from individual electrons. This large current, combined with the large resistance of the tunnel junction, dissipates around $\sim 100\,\mathrm{fJ}$ energy per bit, which is still about two orders of magnitude higher than the switching energy in present day silicon based CMOS technology, and five to six orders larger than the energy barrier for retention. Besides energy dissipation, there is a concern about how to supply the required large number of charges from a transistor of similar size - basically an impedance matching issue. An alternate STT driven nanomagnetic device, the magnetic domain wall based racetrack memory, reduces the resistance by using all metallic structures and separating the read and write paths, but it suffers from even larger current thresholds ($\sim 100\,\mathrm{MA/cm^2}$) because the $\sim 1~\mu m$ sized domains tend to get pinned by notches and defects along the transport pathway \cite{parkin2008magnetic}. 

The efforts to reduce the write current can be summarized into three categories. The first direction involves lowering the critical current required to flip a magnet. In this category, researchers try to dynamically lower the energy barrier that separates different magnetic states. Examples are the adoption of perpendicular magneto-anisotropy (PMA) \cite{ikeda2010perpendicular,nakayama2008spin} and more recently the study of voltage controlled magneto-anisotropy (VCMA) \cite{maruyama2009large,nozaki2010voltage,wang2012electric}. Similar efforts are seen in the domain wall based devices, where researchers are proposing to replace domains with small $\sim 10$ nm sized magnetic skyrmions that need orders of magnitude lower  current threshold ($\sim 100\,\mathrm{A/cm^2}$) \cite{Jonietz2011} for depinning. However, the low current density comes with a price of low operational speed and will need to be exploited judiciously. The second direction is to enhance the charge to spin conversion efficiency for better torque. In usual two terminal MTJs, the spin current is generated by polarizing the electrons that flow through the junction. The spin to charge ratio $I_s/I_q$  normally cannot exceed $100\%$. One way to break this ceiling is to exploit the Giant Spin Hall Effect (GSHE) based on strong spin-orbit coupling in heavy metals like Ta, Pt, W or topological insulators (TI). When an in-plane charge current flows on a metallic film, the spin-orbit coupling generates a spin current that flows perpendicular to the charge current. The spin hall angle $\theta_\mathrm{SHE}=J_s/J_q$ characterizes the intrinsic conversion between spin and charge in those systems, which usually ranges from 0.08 to 0.3 \cite{liu2011spin,Liu2012,pai2012spin}. Placing an MTJ on top of the heavy metal provides an added gain $I_s/I_q=\theta_\mathrm{SHE}L/t>1$ with a geometrical factor $L/t\gg 1$ representing multiple incidences along the skipping orbits over the length of the target soft magnet in the MTJ ($t$ is the thickness of the metal film and $L$ is the length of the magnet along the direction of the charge current). A similar setup can be used on 3D topological insulators like $\mathrm{Bi_2Se_3}$, which have potentially higher $\theta_\mathrm{SHE}\in [2.0,3.5]$ \cite{mellnik2014spin} for surface conduction and may be enhanced by using gated P-N junctions \cite{habib2015chiral}. Finally the third approach is to control the magnetization entirely through an applied electric field, minimizing the energy dissipation that would arise with any steady flow of charge current. Examples here are strain based switching in multiferroic systems. Recent research has experimentally demonstrated magnetization switching in both single phase multiferroic materials \cite{cherifi2014electric,heron2014deterministic} as well as composite structures \cite{d2016experimental}. A winning combination could well involve a hybrid switching scheme combining multiple of these methods, such as mixing the directionality of spin torque with the energetics of strain. The various switching mechanisms will be reviewed in detail in section~\ref{sec:emerging_device}.

\subsection{The Reliability perspective - the high cost of accuracy}
Both memory and logic applications need reliable switching to different degrees. For readout, the read current $I_\mathrm{read}$ needs to be kept small to prevent an accidental switching due to the read disturb. For a write operation, apart from the device-to-device variations, conventional STT switching suffers from a problem with `stagnation'. When the two magnetic layers have precisely collinear (parallel or anti-parallel) magnetic moments, their torque vanishes, whereupon we rely on slow thermal fluctuations to kick the magnetic moment out of this stagnation, which is thus a major source of soft write error. To maintain ultra-low write error rates, much larger overdrive currents are needed beyond the critical current $I_c$ for destabilizing the spins. However, large write currents dissipate more energy and pose a risk of time-dependent dielectric breakdown for MTJs. High switching currents can also lead to MTJ deterioration due to the diffusion of species in the stack. Emerging switching schemes such as VCMA and straintronics can avoid large current (or no current at all) but their write error rates are still too high for practical memory/logic devices. For memory applications, the nanomagnets also need enough energy barrier ($\Delta > 40 k_BT$) to meet the 10 year retention time target. Maintaining enough anisotropy in ultrascaled magnets is hard and requires precise material and interface engineering.

\section{Multi-scale approach to nano-magnetic applications}

\begin{figure}[ht]
\centering
\includegraphics[width=8.5cm]{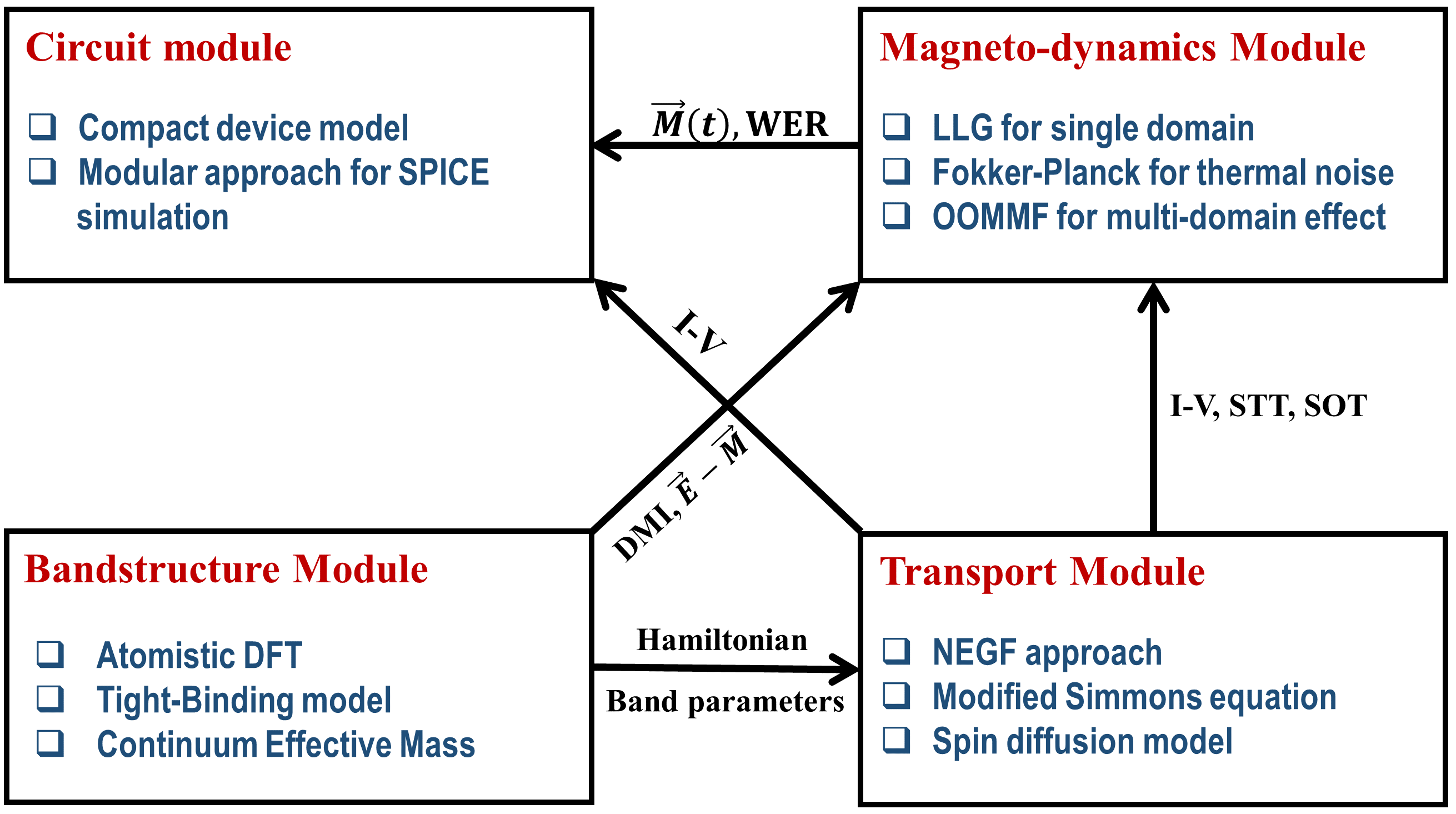}
\caption{Schematic diagram for the multiscale approach to nanomagnetics applications.}
\label{fig:multiscale}
\end{figure}

To understand and resolve these challenges, a holistic approach is needed that goes all the way from the material physics to the circuit performance. In an earlier article \cite{xie2016spin} we laid the groundwork, basic equations, derivations and key concepts underlying this approach. Our focus there was on the physics behind switching - the torque and its voltage asymmetry, symmetry filtering by MgO, the resulting high TMR and its depolarization by oxygen vacancies, and `first principles' spin torques based on Density Functional Theory (DFT) for various materials. The aim of this review article is to focus on the device performance end - i.e., evaluate a few proposals to mitigate challenges at every level, from material to device to circuit. 

Fig.\ref{fig:multiscale} illustrates the multi-scale computational approach. At the material level, we describe the electronic bandstructure of the magnetic contact using atomistic density functional theory (DFT), or a phenomenological continuum model with well benchmarked parameters. These bandstructures allow us to navigate a large phase space of unexplored magnetic materials beyond a few mature material systems, sometimes with opposite design criteria. For example, STT devices need a high spin polarization, low damping and low saturation magnetization for efficient current induced switching, and a high magneto-anisotropy to maintain adequate energy barrier for retention. For voltage driven switching in multiferroic materials on the other hand, we need a low anisotropy field to write easily along the hard axis, and a correspondingly high saturation magnetization to restore the energy barrier. In section \ref{sec:heusler}, we will discuss an example of material engineering involving a high throughput computational study on the Heusler compounds to identify materials with high spin polarization, and layered Heuslers to engineer added magnetic anisotropy. 

The transport module calculates the current-voltage (I-V) characteristics and the spin transfer torque from the magnet and oxide bandstructures - atomistic or continuum. Ballistic transport in MTJs can be calculated using the Non-Equilibrium Green's Function (NEGF) method (section \ref{sec:transport}) \cite{ghosh2015nanoelectronics}, whereas spin transport across larger films such as GSHE or Topological Insulator surface states is described by spin diffusion models. The spin torque is then fed into a magnetodynamics module while the current-voltage (I-V) can be parameterized and used in circuit simulations. Some emerging devices such as those based on multiferroics or skyrmions in magnetic insulators\cite{White2012}, may not need a charge current.  In those cases, the coupling between the ferromagnetic and ferroelectric orders can be extracted from material modeling and directly fed into the micromagnetic simulations. 

At the magneto-dynamic level, the stochastic Landau Lifshitz Gilbert Slonczewski (LLGS) equation describes the dynamics of small magnets in the presence of current driven torque and thermal noise, while the corresponding Fokker-Planck equation directly calculates the nonequilibrium spin probability distribution. Further complications can arise from transient dynamics that may involve incoherent switching with sub-volume excitations. We discuss stochastic magnetization switching in section \ref{sec:reliability}. 

Finally at the highest level, simple compact models (section \ref{sec:analy_modular}), or more general modular circuit models (section \ref{sec:circuit_modular}) incorporating material parameters can be used to simulate a complex architecture and study the interactions among various devices in the circuit. Such a modular approach relies on Kirchhoff's laws for charge current and their relations with the non-conservative spin currents. Ultimately, the merit of this unified toolbox is that it connects the material study all the way to the device/circuit performance metrics approaching a system level analysis, or vice versa - reverse engineering from desired performance criteria back to material optimization.

\section{Material optimization - searching for half-metals in the Heusler family}
\label{sec:heusler}

Although ultra-high TMR was observed under laboratory conditions, the need to reduce resistance area (RA) for impedance matching, and therefore the insulator thickness, reduces the TMR dramatically in practical applications. The development of novel materials is critical to attaining a high TMR ratio beyond $100\%$. This requires a highly spin-polarized magnetic layer in an MTJ. One type of material predicted to have high spin-polarization is a 'half-metal' in which the electronic structure of one of the spin channels is metallic while the other is semiconducting. For Fe and FeCo, this happens to be the $\Delta_1$ band with orbital symmetry given by s, p$_z$ and d$_{3z^2-r^2}$ along the transport z direction. Thus an extremely large TMR is expected to occur in MTJ using half-metals as electrode materials, coupled with MgO which filters out other non-$\Delta_1$ bands through orbital symmetry. The high TMR follows from the fact  that electron tunneling is significantly reduced in the anti-parallel magnetization configuration and reliant on minority spin tunneling.

Although a large number of half-metals have been predicted by first-principles calculations \cite{PhysRevB.66.134428,PhysRevB.51.10436,tobola2000electronic,kandpal2006covalent,galanakis2006electronic}, these discoveries appear to occur mainly through serendipity. Furthermore, it is not clear which of the many half-metals that can be imagined, are stable. Thus, a systematic study of the structural stability of half-metals should provide guidance for future experiments. Here we describe a rational approach to design body-center cubic (BCC) half-metallic structures called ``Heusler compounds"\cite{butler2011rational}. We also investigate the stability of these compounds through the calculation of their formation energies and the comparison of these calculated energies to the calculated energies of other possible phases, and combinations of phases \cite{PhysRevB.95.024411}.

\subsection{How to design a half-metal}

\begin{figure}
\centering
\includegraphics[width=\columnwidth]{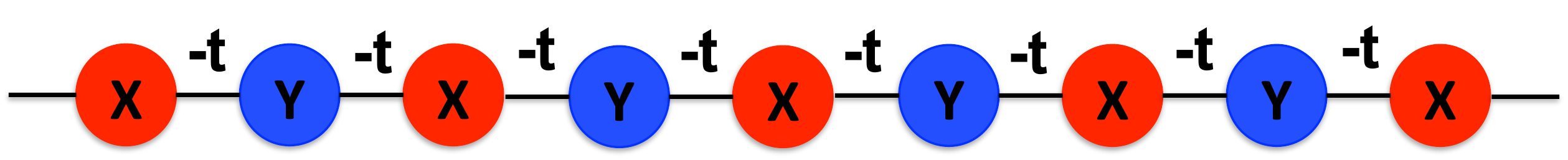}
\caption{1D atomic system with two sub lattices, type $X$ atom sits on one sub-lattice and type $Y$ atom sits on the other. There is only nearest neighbor interaction and the interaction is represented by the hopping parameter $-t$}
\centering
\label{fig:1Dchain}
\end{figure}

To explain the origin of half-metals in the Heusler family, we need to start from a 1D atomic system with two different atoms, $X$ and $Y$, as shown in Fig.\ref{fig:1Dchain}. We assume different on-site energies of the orbitals for $X$ sites and $Y$ sites, $E_{X}$ and $E_{Y}$ ($E_{X}<E_{Y}$), respectively. If we only consider the first-nearest-neighbor interaction, the 1D atomic chain can be described by the tight-binding Hamiltonian for the periodic array of dimers:

\begin{equation}
\begin{aligned}
E\{\psi\}_{n}&=[\alpha]\{\psi\}_{n}-[\beta]\{\psi\}_{n+1}-[\beta]^{\dagger}\{\psi\}_{n-1}\\
[\alpha]&=\left[\begin{array}{cc}
E_{X} & -t\\
-t & E_{Y}
\end{array}\right],[\beta]=\left[\begin{array}{cc}
0 & 0\\
t & 0
\end{array}\right]
\end{aligned}
\label{eq:1Dchain}
\end{equation}

\begin{figure}
\centering
\includegraphics[width=\columnwidth]{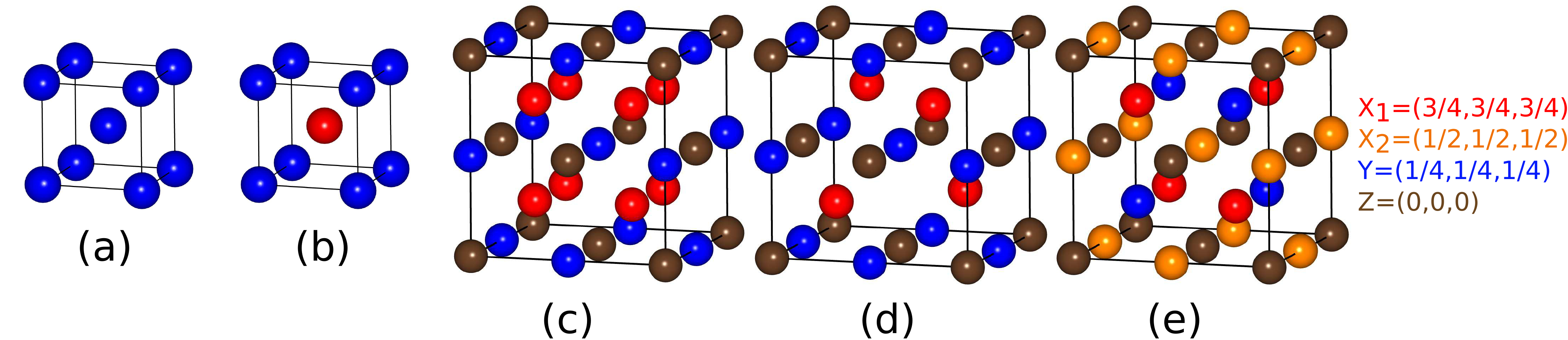}
\caption{Schematic representation of full-Heusler L2$_{1}$ structure, half-Heusler C1$_{b}$ structure and inverse-Heusler XA structure. (a) The BCC body-center cubic structure. (b) The B2 cubic structure. (c) The L2$_{1}$ structure consists of four interpenetrating fcc sublattices with atomic sites X$_{1}$ $\left(\frac{3}{4},\, \frac{3}{4},\, \frac{3}{4}\right)$,  X$_{2}$ $\left(\frac{1}{4},\, \frac{1}{4},\, \frac{1}{4}\right)$, Y $\left(\frac{1}{2},\, \frac{1}{2},\, \frac{1}{2}\right)$, and Z $(0,\, 0,\, 0)$. (d) The C1$_{b}$ structure consists of three interpenetrating fcc sublattices with atomic sites X $\left(\frac{1}{4},\, \frac{1}{4},\, \frac{1}{4}\right)$, Y $\left(\frac{1}{2},\, \frac{1}{2},\, \frac{1}{2}\right)$, and Z $(0,\, 0,\, 0)$. (e) The XA structure consists of four interpenetrating fcc sublattices with atomic sites X$_{1}$ $\left(\frac{3}{4},\, \frac{3}{4},\, \frac{3}{4}\right)$,  X$_{2}$ $\left(\frac{1}{2},\, \frac{1}{2},\, \frac{1}{2}\right)$, Y $\left(\frac{1}{4},\, \frac{1}{4},\, \frac{1}{4}\right)$, and Z $(0,\, 0,\, 0)$. In the XA structure, X$_{1}$ and X$_{2}$ have the same transition metal elements and their valence smaller than Y.}
\label{fig:B2toInverse}
\end{figure}

\noindent where $-t$ is the coupling between nearest neighbor atoms. 
Assuming periodic boundary conditions and invoking a plane
wave solution (Bloch's theorem) $\psi_{n}\sim e^{ikna}$, we can get a band gap extending from $E_{X}$ to $E_{Y}$ with bandgap $E_g=E_{Y}-E_{X}$.
This gap is independent of the hopping term, $-t$, letting us easily extend the 1D chain to 3D body-center cubic B2 structures\cite{PhysRev.94.1498,butler2011rational}. B2 structure can be viewed as alternate atomic layers along (001) direction only with nearest neighbor interactions (As shown in Fig. \ref{fig:B2toInverse}(b)), but there are no intra-layer interactions. This means that the wave vector perpendicular to the (001) plane, $k_{\perp}$,  generates an $E-k$ expression identical to the 1-D dimer $E-k$ expression above, resulting in a gap between $E_{X}$ and $E_{Y}$. Since the gap between $E_{X}$ and $E_{Y}$ is not dependent on $k_{\parallel}$, the gap will remain after integration over $k_{\parallel}$.

In order to create half-metallic ferromagnets, we can select different transition metal elements with $d-$orbitals on the $X$ and $Y$ sites. As long as the interactions are restricted to nearest neighbors, we can predict that there is still a gap between the onsite energy for the $X$ sites and the onsite energy for the $Y$ sites for a B2 structure.

If we assume the onsite energy is determined primarily by the onsite electron number, then we see that the system can lower its energy if it enhances the size of the gap by increasing the difference in the number of electrons for $X$ and $Y$ atoms, so that the Fermi energy falls inside the gap. The outermost electronic states that participate in hybridization in B2 compounds are a single $4s$ and five $3d$ orbitals, so we have nearly six bonding states below the Fermi energy. In other words, there are a total of six electrons in the gapped spin channel. We thus end up with an approximate thumb rule for these compounds: in the gapped spin channel, the approximate electron count on alternate atoms $X$ and $Y$ must be four and two (the `4-2' rule) \cite{butler2011rational}. 
If the total valence electron number is twelve, it is easy then to create a semiconductor since each spin channel has six electrons. By making the total valence electron number larger (or smaller) than 12, the majority (or minority) spin channel will now be 'doped' to be metallic, while the other spin channel remains semiconducting since the Fermi energy falls in the gap.


While the above design rule seems compelling, first principles calculations indicate that the B2 structure only achieves a pseudogap instead of a true gap. This is because of residual next nearest neighbor (NNN) interactions \cite{butler2011rational} that were not included in our consideration so far. To reduce the NNN interactions, we can replace one transition $Y$ atom with a main group $Z$ atom without any d-states, acting as a `spacer' - resulting in the full-Heusler family of compounds $X_{2}YZ$ (Fig.\ref{fig:B2toInverse} c). Furthermore, We can delete half of the $X$ atoms to generate half-Heusler compounds $XYZ$ with a `cleaner' gap (Fig.\ref{fig:B2toInverse} d). We can also exchange the position of $Y$ atoms and $X$ atoms to generate inverse-Heusler compounds $X_{2}YZ$ (Fig.\ref{fig:B2toInverse} d). In all cases, the decreased NNN interactions will narrow the bands and prevent the density of states from tailing into the gap.

\begin{figure}[t]
\centering
\includegraphics[width=\columnwidth]{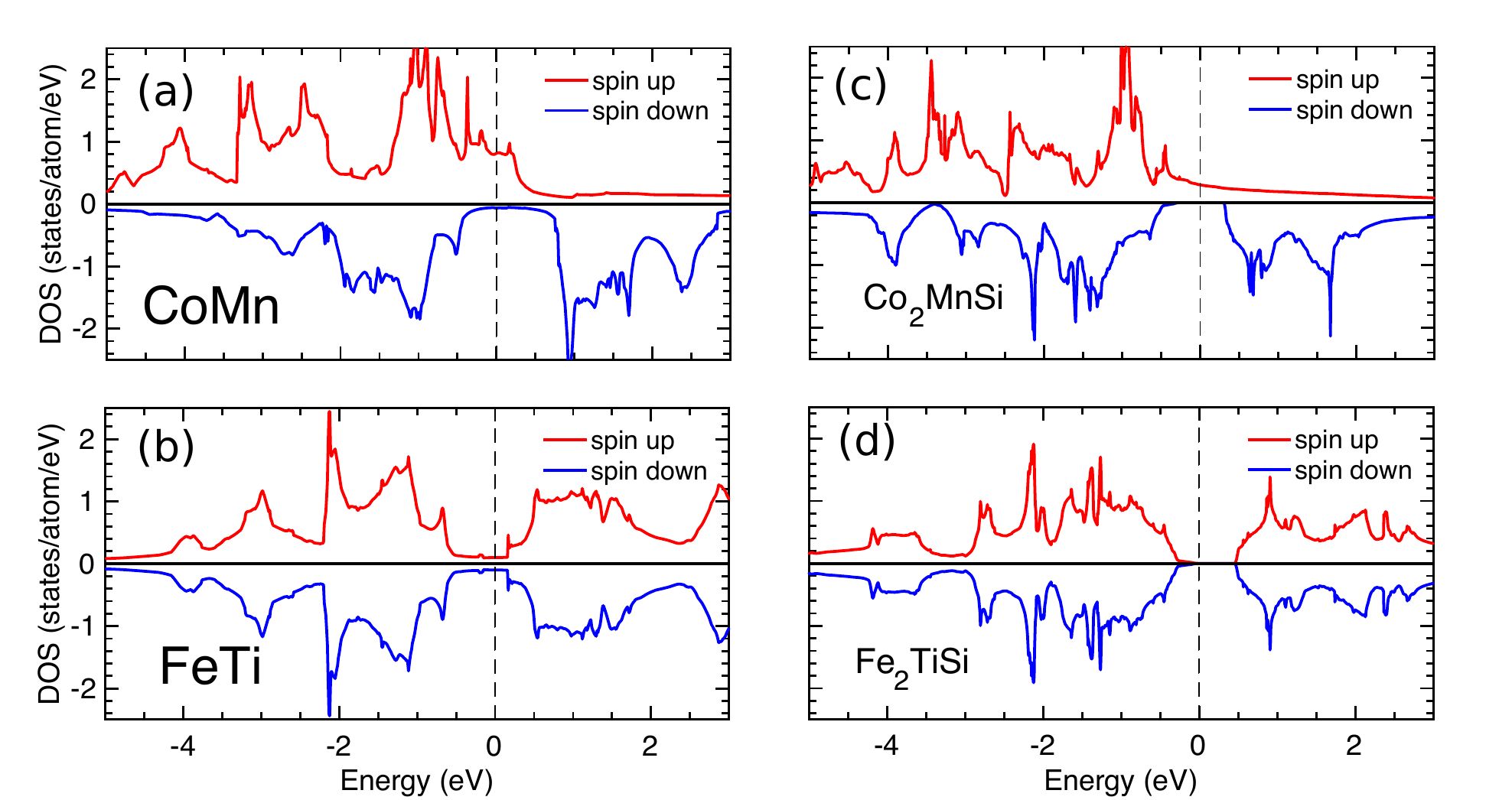}
\caption{Density of States (DOS) for (a) CoMn, (b) FeTi, (c) Co$_2$MnSi, and (d) Fe$_2$TiSi. The blue color is for the majority spin channel and the red color is for the
minority spin channel.}
\centering
\label{fig:CoMnFeTi}
\end{figure}

To confirm this rational design approach, We performed all calculations using Density-Functional Theory (DFT) as implemented in the Vienna Ab-initio Simulation Package (VASP)~\cite{Kresse199615} to calculate the density of states for CoMn and FeTi in the B2 phase. Rather than a gap, we find a deep, wide minimum in the minority spin density of states at half-filling which we refer to as a pseudogap, with 3 electrons per atom under the Fermi energy. For the pseudogapped channel, the “4-2 rule” forces $X$ atoms (Fe or Co) to hold 4 electrons and $Y$ atoms hold 2 electrons. As we replace one Mn or Ti with a Si spacer, the pseudogaps in B2 systems turn into real gaps, resulting in the half-metal Co$_{2}$MnSi and semiconducting Fe$_{2}$TiSi. For these two full-Heusler compounds, the spacer Si atom also holds 2 electrons in the gapped channels to sustain the `4-2' rule. We can thus create half-metals by turning B2 compounds into Heusler compounds.

Based on this idea, a large number of layered half-metallic structures can be obtained in full-Heusler, half-Heusler, and inverse-Heusler family. Let us discuss half-Heusler compounds in the next section.

\subsection{Half-Heusler compounds}

In this section, we review our computational investigation covering 378 half-Heuslers (with $X=$ Cr, Mn, Fe, Co, Ni, Ru, Rh;  $Y=$ Ti, V, Cr, Mn, Fe, Ni;  $Z=$ Al, Ga, In, Si, Ge, Sn, P, As, Sb) as calculated by VASP\cite{PhysRevB.95.024411}.  The crystal structure of half-Heuslers is shown in Fig.\ref{fig:B2toInverse} (d). 

\begin{figure}[b]
\centering
\includegraphics[width=\columnwidth]{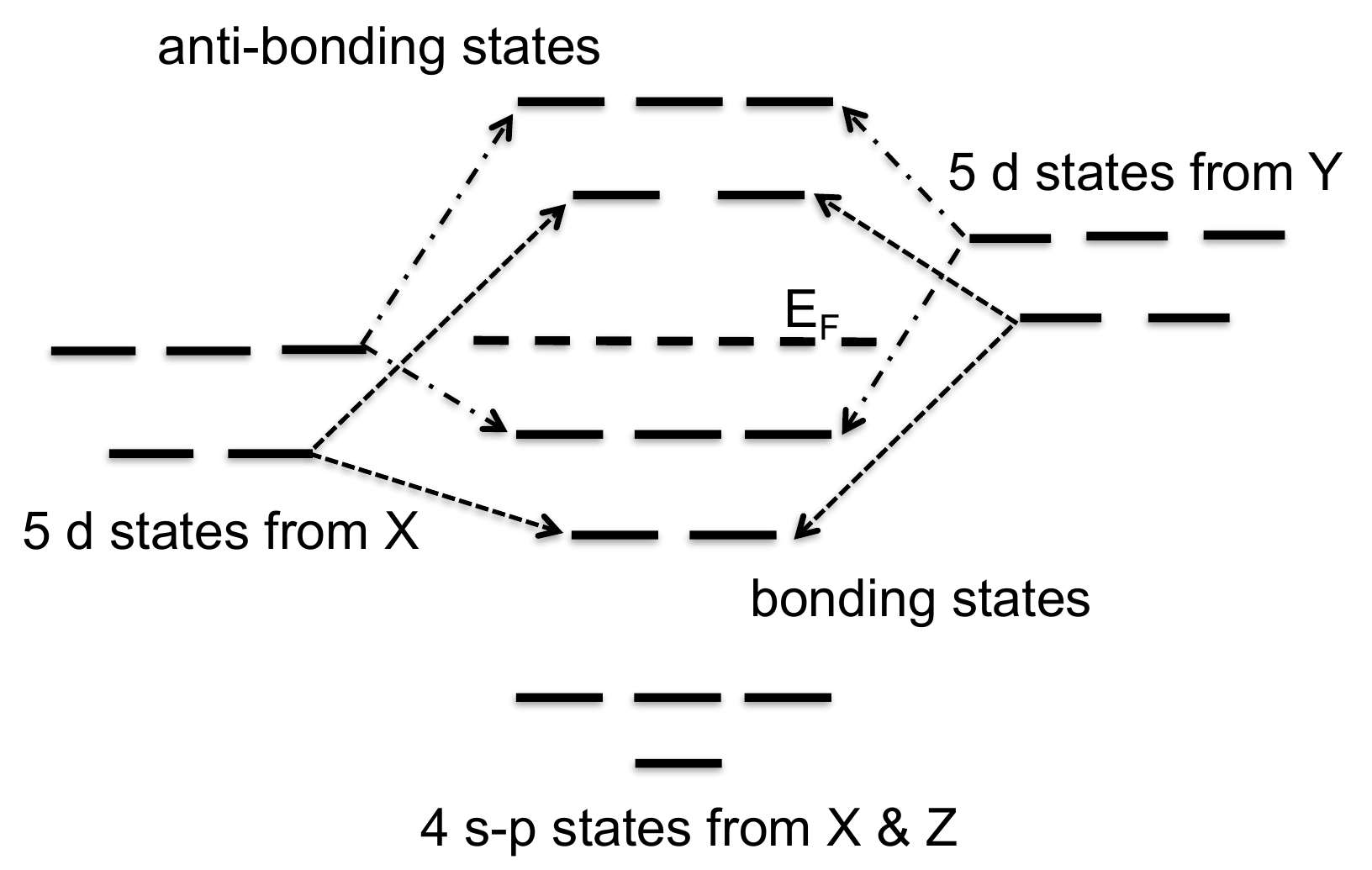}
\caption{Schematic illustration of hybridization in the gapped channel of the half-Heusler compounds: The energy levels of the energetically lower lying bonding $d$ states are separated from the energy levels of the anti-bonding $d$ states by a gap, such that only the 5 bonding $d$ states are occupied. The lowest occupied 4 $s-p$ states come from the hybridization between $X$ and $Z$ elements.}
\centering
\label{fig:18states}
\end{figure}

If we consider upto NNN interactions, the nearest-neighbor couplings will only occur between $Y-X$ and $Z-X$ atoms. Since the four $s-p$ orbitals from the main group element $Z$ hybridize with a lower energy level, the origin of the gap in the gapped spin channel comes from the hybridization between the five $3d$ states of the higher valence and the lower valence transition metal atoms $X$ and $Y$. As the Fermi energy falls in the gap, there are a total of $4+5=9$ states being filled in this spin channel (as shown in Fig.\ref{fig:18states}). Since the spin moment per atom is just the difference in the number of up and down electrons per atom ($M=N^\uparrow -N^\downarrow$), and since $N=N^\uparrow + N^\downarrow$, with 9 states fully occupied in the gapped channel  ($N^\downarrow=9$) we get the simple `rule of 18 for half-metallicity in half-Heusler compounds:
\begin{equation}\label{eqn:SP_rule}
M_{tot}=N_{V}-18
\end{equation}
where $M_{tot}$ is the total magnetic moment and $N_{V}$ is the total number of valence electrons per \textit{XYZ} formula unit. If the total valence electron number is 18, the system will tend to be semiconducting with 9 electrons under the Fermi energy per spin channel. If the total valence electron number is not equal to be 18, one spin channel may be 'doped' to be metallic while the other spin channel remains semiconducting, resulting in a half-metal. We find from our computational results that all the semiconductors and half-metals in the half-Heusler family do follow this simple Slater-Pauling rule precisely with integer total moments.
\begin{figure}
\centering
\includegraphics[width=\columnwidth]{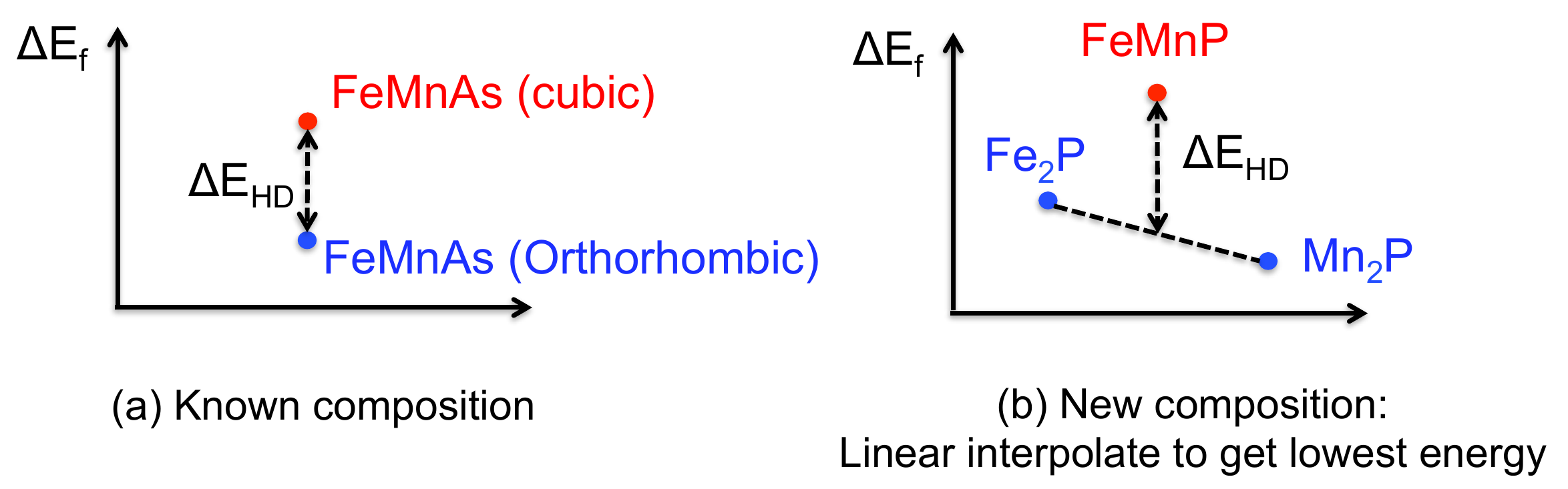}
\caption{(a) Hull distance $\Delta E_{\rm HD}$ for half-Heusler FeMnAs (b) Hull distance $\Delta E_{\rm HD}$ for half-Heusler FeMnP}
\centering
\label{fig:hull_distance}
\end{figure}

We further investigate the energetics of 378 half-Heusler compounds in the cubic structure to ensure their stability. For each \textit{XYZ} half-Heusler compound, we calculate its formation energy $\Delta E_{f}$ using Eq.~\ref{eqn:formation_energy} and distance from the convex hull $\Delta E_{\rm HD}$ using Eq.~\ref{eqn:hull_distance}. 
\begin{equation}\label{eqn:formation_energy}
\Delta E_{f}\,(XYZ) = E\,(XYZ) - \frac{1}{3}\left(\mu_X + \mu_Y + \mu_Z\right)
\end{equation}

\begin{equation}\label{eqn:hull_distance}
\Delta E_{\rm HD} = \Delta E_{f} - E_{hull}
\end{equation}

\noindent where $E(XYZ)$ is the total energy per atom of the half-Heusler compound, $\mu_i$ is the reference chemical potential of element $i$, chosen to be consistent with those used in the OQMD database (See Ref.~\cite{oqmd_npj_2015} for details), and $E_{hull}$ is the lowest formation energy at the composition of all the known phases. A negative value of $\Delta E_{f}$ indicates that at zero temperature, the half-Heusler compound is more stable than its constituent elements.  However, it does not guarantee the stability of a half-Heusler phase over another competing phase or a mixture of phases, which is where the Hull distance comes in. 


A measure of the thermodynamic stability of a phase is its distance from the convex hull. We utilize the formation energy data from OQMD to do the phase competing analysis \cite{oqmd_npj_2015,ADMA:ADMA200700843,AENM:AENM201200593}. In Fig.\ref{fig:hull_distance} (a), the hull distance $\Delta E_{\rm HD}$ for cubic FeMnAs is the formation energy difference between the cubic phase and the orthorhombic phase, which has the lowest formation energy at that chemical composition. In Fig.\ref{fig:hull_distance} (b), although we don't have available the formation energy of other phases at that chemical composition, we can still  estimate stability using the hull distance $\Delta E_{\rm HD}$ for cubic FeMnP by defining the hull energy ($E_{hull}(FeMnP)$) as a linear combination of formation energies of nearby phases Mn$_{2}$P and Fe$_{2}$P.  By comparing with reports for hull distances of experimentally synthesized ternary compounds, we conclude that a Heusler compound with $\Delta E_{\rm HD}<$ 0.1 eV has a strong likelihood of experimental synthesis \cite{PhysRevB.95.024411}.

\begin{figure*}
\centering
\includegraphics[width=5in]{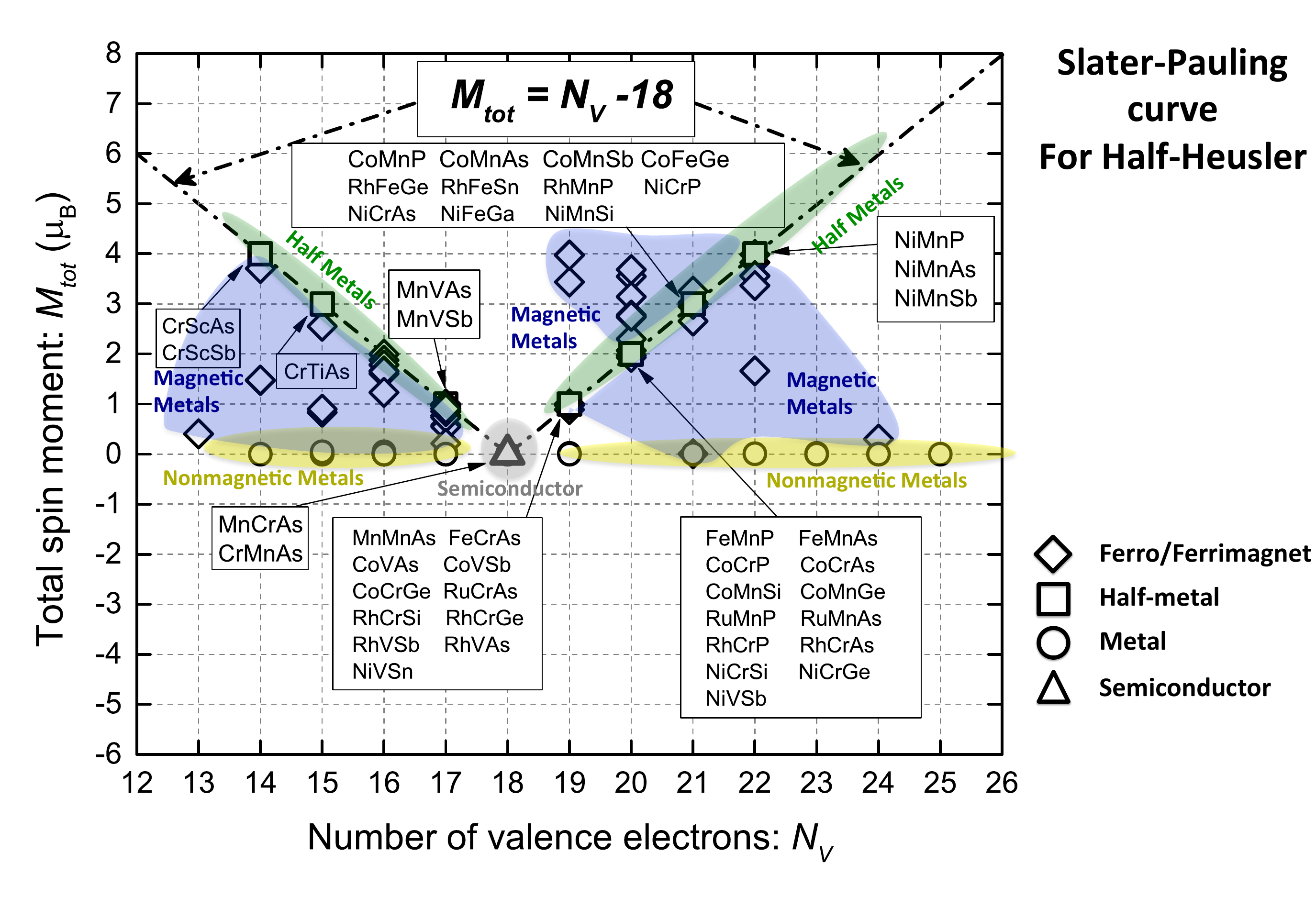}
\caption{Calculated total magnetic moment $M_{tot}$ as a function of the total number of valence electrons $N_V$ per formula unit for the 203 half-Heusler compounds with negative formation energies. The dash-dot line represents the Slater-Pauling rule $M_{tot}=N_{V}-18$, and all the 45 half-metals and 26 semiconductors follow this rule precisely. Diamond, square, circle, and triangle symbols indicate ferro/ferrimagnets, half-metals, metals, and semiconductors, respectively. We also use different color shapes to mark off half-metals, semiconductors, and magnetic/nonmagnetic metals. To avoid confusion about the signs of magnetic moments, we uniformly use the absolute values of magnetic moments in this diagram.}
\label{fig:spin_moment_vs_valence_electrons}
\end{figure*}

From our calculations, we discovered 26 semiconductors and 45 half-metals with negative formation energy. All the half-metals (listed in boxes and square symbols in the green ovals surrounding the dashed lines) and semiconductors (triangles in the gray circle at the base of the dashed lines) are summarized in Fig.\ref{fig:spin_moment_vs_valence_electrons} and all follow the Slater-Pauling rule (Eq.\ref{eqn:SP_rule}) precisely. All the other magnetic/nonmagnetic metals are marked out in the purple and yellow areas. 17 out of the 26 semiconductors are identified with $\Delta E_{\rm HD}<$0.1 eV, including 3 likely candidates for fabrication (CoVSn, RuVAs, and RhVGe) that have not been experimentally reported so far. Furthermore, 11 half-metals are identified with $\Delta E_{\rm HD}<$ 0.1 eV.  9 of these have already been experimentally synthesized in bulk, albeit not in a half-Heusler phase, but they may be synthesized in thin films by molecular-beam epitaxy (MBE). Among the two remaining half-metals, NiMnSb has been verified as a half-Heulser compound with half-metallic characteristics in bulk\cite{PhysRevB.42.1533}. Although CoVSb has also been well established in a cubic phase, the compound is a weak itinerant ferromagnet rather than a half-metal\cite{CoVSbMag,CoVSbNeutron}. As a result, only one half-metal RhVSb with 
 $\Delta E_{\rm HD}= $0.103 eV can be predicted as a good candidate for future experiments. 

While the half-Heusler family ultimately produced few stable half-metals, in hindsight this is not hard to understand because the XYZ cubic structures are not close packed and contain a vacancy sublattice, which compromises their stability. Based on our studies, we expect the full-Heusler family to yield a larger number of stable half-metals.

\subsection{Non-magnetic spacer engineering - symmetry filtering of MgO}

\begin{figure}
\centering
\includegraphics[width=8.5cm]{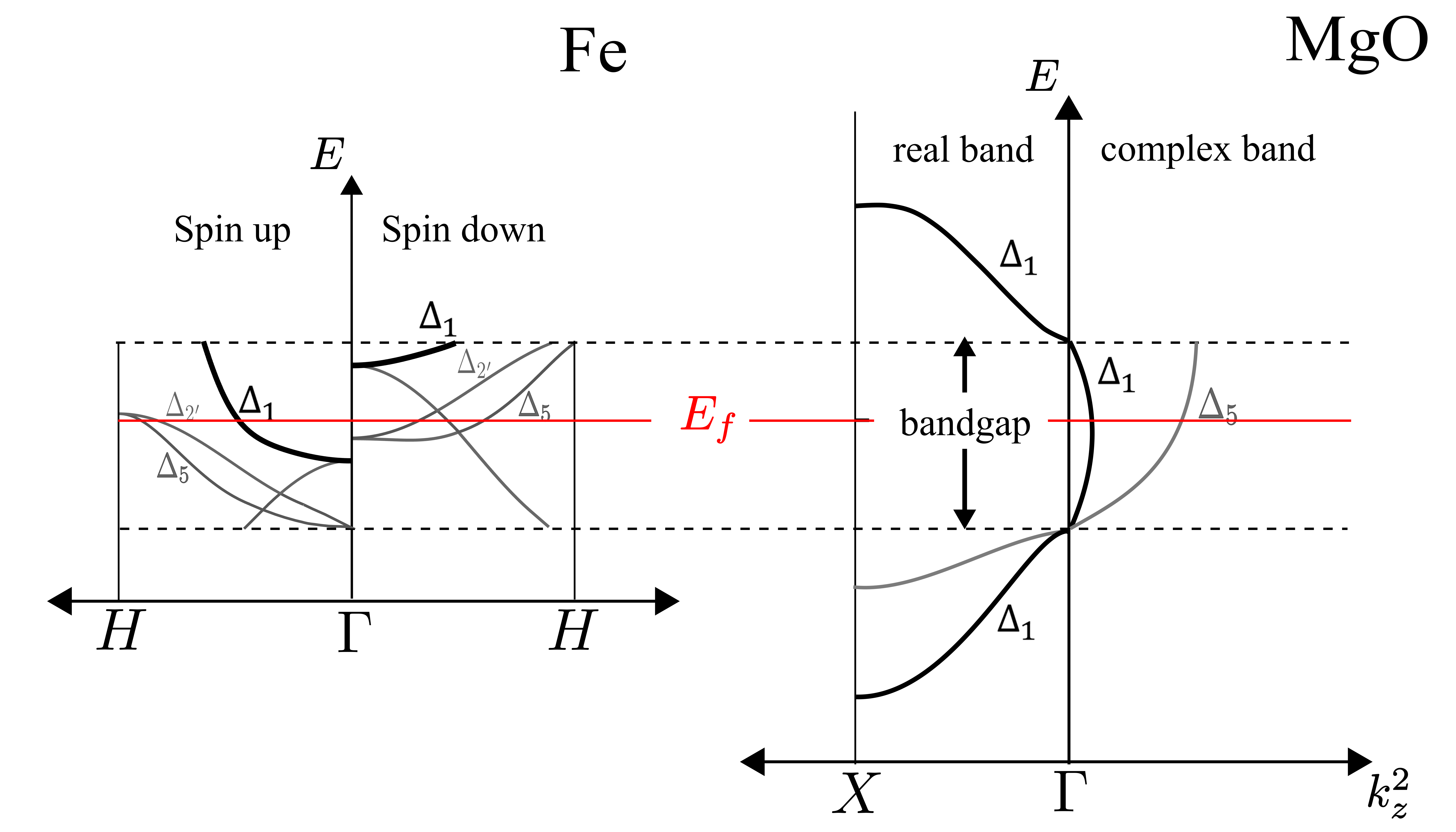}
\caption{Bandstructure of bcc Fe and MgO. For Fe, the bands are plotted along the transport direction for different spin channels. For MgO, the real band is plotted on the left side and the complex band inside the bandgap (from ref.\cite{butler2001spin}) is plotted on the right side. Bands with different symmetries $\Delta_1,\Delta_5,\Delta_{2'}$ are indicated. The weakest decaying band is the $\Delta_1$ band connecting the conduction and valence bands, which also happens to be half-metallic in Fe (only the up spin crosses the Fermi energy). The dual effect of symmetry filtering and half-metallicity makes the predicted ballistic TMR of Fe/MgO a record high ($> 4000 \%$)}
\centering
\label{fig:MgO_symmetry}
\end{figure}

The other important material choice for the magnetic tunnel junction is the nonmagnetic spacer. One of the important developments in MTJ technology was the adoption of MgO as the spacer material in magnetic tunnel junctions. As predicted in Fe/MgO/Fe junction from the first-principle calculations\cite{butler2001spin}, the reason why MgO performs much better than the common earlier $\mathrm{Al_2O_3}$ as the insular in MTJs is due to the symmetry filtering effect. When the propagating states in Fe hit the MgO, these wavefunctions continue inside the bandgap of MgO as decaying waves. The continuation requires the conservation of symmetry from the Fe states to the MgO states. In MgO, different states have different decaying lengths. In particular, the complex $\Delta_1$ band in MgO needs to connect the conduction and valence $\Delta_1$ bands with similar orbital constitutents, which results in a shorter wavelength (longer decaying length) near the mid-gap than the simple WKB theory predicts. In an Fe/MgO/Fe magnetic tunnel junction, the Fermi energy lies near the mid-gap of MgO, as shown in Fig.\ref{fig:MgO_symmetry}. As a result, other Fe bands are filtered out and leaves the Fe $\Delta_1$ band with the highest transmission through MgO, dominating the tunneling current. As it turns out, the Fermi energy only crosses the Fe $\Delta_1$ band in one spin channel, which essentially makes Fe a half-metal to MgO. A more detailed discussion about symmetry filtering can be found in ref.\cite{butler2001spin,xie2016spin}. Noted that in practical devices CoFe is better than pure Fe because of anisotropy and magnetostriction, but the same symmetry filtering mechanism applies.

Note that MgO is particularly critical to the performance of  MTJs built from non half-metallic magnetic electrodes such as Fe and FeCo, where multiple bands cross the Fermi energy. In these MTJs, symmetry filtering is needed to remove all but the $\Delta_1$ band, which alone needs to be half-metallic. 
For intrinsic half-metals on the other hand such as the Half-Heuslers described earlier, we seek instead to find any insulating spacer that preserves its half-metallicity. MgO might also satisfy that function in some cases, for instance with NiMnSb as we show in section \ref{sec:transport}. In the next section, we will discuss another class of insulators that may offer better opportunities for maintaining the half-metallicity of Heusler half-metals - namely Heusler semiconductors. Combining a Heusler metal with a lattice matched Heusler semiconductor brings in another advantage - creating a uniaxial anisotropy, which is also important in nanomagnetic applications. The combination of different Heusler alloys opens the possibilities of better engineered heterojunctions for spin transport.  In fact, the metallic 'All-Heusler' spin valves have already been investigated for enhanced current-perpendicular-to-plane giant magnetoresistance effect (GMR)\cite{nikolaev2009all,li2016current}.

\subsection{All-Heusler superlattice: half metals with uniaxial anisotropy}

\begin{figure}
\centering
\includegraphics[width=2.8in]{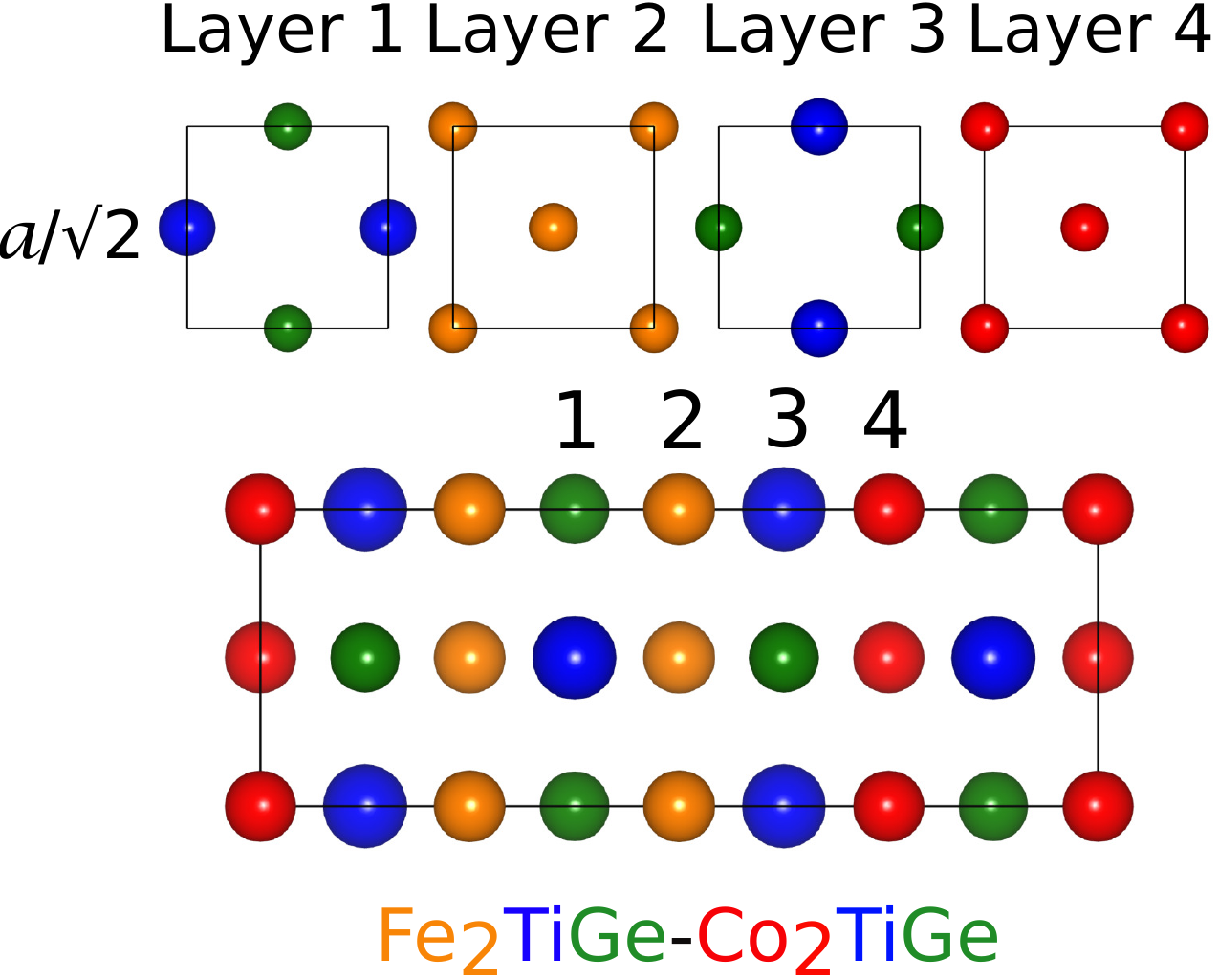}
\caption{Layer by layer depiction of the heterostructure unit cell layered in the [001] direction for the full Heusler superlattice Fe$_2$TiGe-Co$_2$TiGe. The distance between layers is $a$/4.}
\centering
\label{fig:heuslersuperlattice}
\end{figure}

One drawback of Heusler compounds is that the symmetry of their $B_2$ cubic crystal structure compromises their uniaxial magnetocrystalline anisotropy (MCA), needed to create a large enough energy barrier for spin retention. However, our study unearthed several half-Heusler compounds that were not only half-metals, but also semiconductors, many with very similar lattice constants. This is an important factor because half-metallicity arose from the specifics of the underlying crystal structure and chemistry, and extending that crystal structure in a heterojunction helps maintain the half-metallicity of the candidate Heuslers. Indeed from  first-principles calculations, we show that by making layered superlattices of two Heusler compounds, an intrinsic uniaxial anisotropy can be achieved \cite{heuslersuperlattice} while maintaining the half-metallicity.  This uniaxial anisotropy is generated by the different electronic configurations of the two Heusler compounds and by the distortion of the lattice, which causes the local environment of each atom to be different in the direction perpendicular to the layers from that in the plane of the layers. We find that some of the Heuser superlattices layered perpendicular to the [001], [110] and [111] directions do retain their half-metallicity. We find this to be true for full-full, half-half and full-half Heusler combinations. 

In short, some of the designed superlattices combine half-metallicity with perpendicular magnetic anisotropy. For example, the Co$_{2}$TiGe-Fe$_{2}$TiGe superlattice layered along the [100] direction (shown in Fig.\ref{fig:heuslersuperlattice}) can retain half-metallicity with a perpendicular anisotropy of $3.421 \times 10^{5}$ J/m, while the Co$_{2}$MnSi-CoTiSi superlattice layered along the [110] direction retains its half-metallicity with perpendicular anisotropy of $2.699 \times 10^{5}$ J/m. More details are discussed in ref\cite{heuslersuperlattice}. Such PMA half-metallic superlattices or thin films can be quite competitive as a material stack for an STT-MRAM device.
 
\section{Transport in nanostructures with NEGF}
\label{sec:transport}
\subsection{Non-equilibrium Green's Function (NEGF) approach}
\subsubsection*{Charge Current in Nanostructures.} 
NEGF is a powerful tool to study the quantum transport phenomenon in nanostructures such as the magnetic tunnel junctions. In a two-terminal device, the charge current can be evaluated using
\begin{equation}
\begin{split}
I(V)&=\frac{q}{h}\int dE T(E;V)(f_L-f_R) \\[0.5ex]
T&=\mathrm{Tr}[\Gamma_L(E) G^\dagger(E) \Gamma_R(E) G(E)]
\end{split}
\label{eq:IV}
\end{equation}
where $T(E;V)$ is the bias $V$ dependent transmission between the left and right electrodes at a given energy $E$. $f_{L,R}$ are the Fermi functions of the left and right electrodes respectively. $G(E)$ is the retarded Green's function and $\Gamma_{L,R}(E)$ are the electrode coupling matrices given by:
\begin{equation}
\begin{split}
G(E)&=\left[E-H-\Sigma_L(E)-\Sigma_R(E)\right]^{-1}\\
\Gamma_{L,R}(E)&=i\left[\Sigma_{L,R}(E)-\Sigma^\dagger_{L,R}(E)\right]
\end{split}
\end{equation}
where $H$ is the Hamiltonian of the system and $\Sigma_{L,R}(E)$ are the `self-energies' that account for the projected bands and level broadenings by the two semi-infinite electrodes \cite{ghosh2015nanoelectronics,datta2005quantum,rocha2006spin}.
\subsubsection*{Spin Current and spin transfer torque}
One can use the NEGF equations to calculate the spin current and spin transfer torque as well. The spin torque $\boldsymbol{\tau}$ can be calculated from the time evolution of the magnetic moment of electron $\mathbf{M}=\mathrm{Tr}\left<\psi^\dagger\mathbf{S}\psi\right>$, which is derived from the Schrodinger equation, $\partial\psi/\partial t=H\psi/i\hbar$, $\partial\psi^\dagger/\partial t=-\psi^\dagger H/i\hbar$:
\begin{equation}
\begin{split}
\boldsymbol{\tau}&=\frac{\partial\mathbf{M}}{\partial t}=\frac{1}{i\hbar}\mathrm{Tr}
\left< -(\psi^\dagger H)\mathbf{S}\psi+\psi^\dagger\mathbf{S} \left(H\psi\right)\right>\\
&=-\frac{i}{2}\mathrm{Tr}\left[\boldsymbol{\sigma}(G^nH-HG^n)\right]\\
&=-\frac{i}{2}\sum_{j}\boldsymbol{\sigma}\left[G^n_{ij}H_{ji}-H_{ij}G^n_{ji}\right]\\
&=-\sum_j \mathbf{I}^s_{ij}
\end{split}
\end{equation}
where the spin operator $\mathbf{S}=\hbar\boldsymbol{\sigma}/2$ and $\boldsymbol{\sigma}=(\sigma_x,\sigma_y,\sigma_z)$ are the Pauli matrices. $G^n=\langle\psi\psi^\dagger\rangle$ is the electron correlation function describing the occupancy of the states. 
The spin current $\mathbf{I}^s$ between two sites $i$ and $j$ can be extracted from the torque expression as\cite{ghosh2015nanoelectronics,theodonis2006anomalous}:
\begin{equation}
\mathbf{I}^s_{ij}=\frac{i}{2}\boldsymbol{\sigma}\left[G^n_{ij}H_{ji}-H_{ij}G^n_{ji}\right]
\label{eq:spin_current}
\end{equation}
The above equation works for systems with orthogonal basis. For non-orthogonal basis, an additional overlap matrix should be included in the formalism. Details can be found in \cite{xie2016spin}. The calculated spin current can then be used to evaluate the spin transfer torque based on the angular momentum conservation. For example, to evaluate the STT on the free magnetic layer of the MTJs, we calculate the incoming spin current at the oxide-magnet interface. We can then assume the spin current coming out from the other side is already fully polarized along the magnetic moment of the free layer, implying that all the angular momentum perpendicular to the free layer has been absorbed, which equals the spin torque exerted on the free magnet.

\subsection{{\it ab-initio transport calculations}}
Coupling NEGF with DFT allows parameter-free first principles calculation of the I-V characteristics and the spin transfer torque of the magnetic tunnel junction. As previously discussed, the high throughput computational studies of Heusler compounds have provided us with a large pool of potential materials for nanomagnetic applications. These half-metals are critical to build effective MTJs with high TMR ratio. To investigate whether their half-metallicity can be preserved in an MTJ configuration, and to quantify how much STT and TMR a Heusler-based MTJ or an all-Heusler superlattice can generate, we need to move from electronic structure to nonequilibrium transport calculations. For an MTJ structure for instance, we can use LLGS to get the macrospin dynamics in presence of a drive current, but to relate that current back to an applied voltage, we will need to calculate the current-voltage (I-V), with barrier heights, contact densities of states and orbital effective masses of the tunneling electrons extracted from DFT. We use the numerical atomic orbital-based DFT program {\it SIESTA} that combines DFT with the Non-equilibrium Green's Function program {\it Smeagol} to calculate the I-V characteristics and the spin transfer torque of a magnetic tunnel junction from first principles \cite{xie2016spin,rocha2006spin,soler2002siesta}.
\begin{figure}[ht]
\centering
\includegraphics[width=8.5cm]{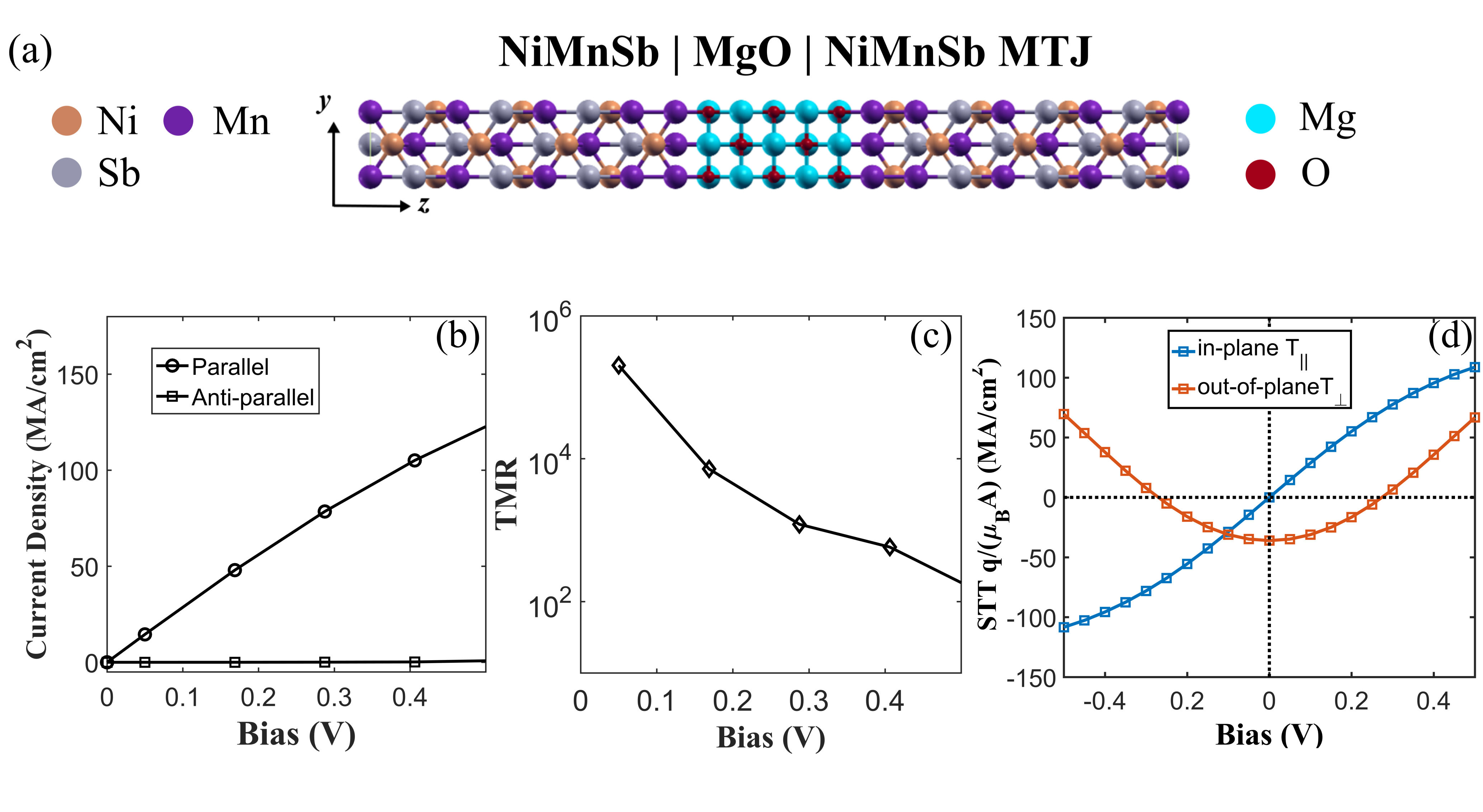}
\caption{Transport calculation for a NiMnSb-MgO-NiMnSb junction. (a) Unit cell used for NiMnSb-MgO-NiMnSb heterojunction with 5 monolayers of MgO. (b) The low bias I-V characteristics. (c) Low bias TMR ratio calculated from the I-V: $\mathrm{TMR}=(I_p-I_{ap})/I_{ap}$. (d) The bias-dependent spin transfer torque of the junction when the magnetic moments on the two sides are at $\pi/2$. }
\label{fig:IV_STT}
\end{figure}

Fig.\ref{fig:IV_STT} shows an example calculation on the half-meta\-llic NiMnSb-MgO(5ML)-NiMnSb junction. The interfacial structure is chosen from the lowest energy configurations with the inter-atomic distance averaged from a fully relaxed structure. The ultra-high TMR ($>10^4$) indicates that the {\it{half-metallicity is well preserved in an MTJ even in the presence of interfacial strain and applied bias}}. The calculated TMR is most likely overestimated because of the adoption of a ballistic transport model across the MgO (predominantly driven by tunneling). We do not include any spin-flip scattering in our current calculations at this time. Since $I_{ap}$ is a small number, a tiny amount of scattering could vastly degrade the TMR. In principle, the defect induced spin scattering can be included in the calculation, but that requires prior knowledge of the type and location of depolarizing defects in the MTJ and the exact interface morphology. For example, we reported the impact of oxygen vacancies on TMR degradation in Fe/MgO/Fe MTJs in \cite{xie2016spin}. We found there that the TMR is particularly compromised if the vacancy sits close to the oxide-magnet interface. We can include incoherent spin scattering in transport calculations by representing the ensemble of the scattering events with a self-energy term in NEGF in the Born approximation, describing their average effect on the transport \cite{Yanik2007}. This method however needs prior microscopic knowledge of the spin scattering potential at a given interface morphology. 

\subsection{Spin amplification in Topological Insulators}
High spin-polarized magnetic electrode materials are critical to the improvement of the TMR ratio and the readability of nanomagnetic devices. Increase in spin polarization also improves the write performance in STT-MRAMs. However, without a spin-to-charge gain, it is hard to reduce the write current beyond a limit set by conservation of spin angular momentum between the conduction electrons and the magnetization of the electrode.  
\begin{figure}[ht]
\centering
\includegraphics[width=8.5cm]{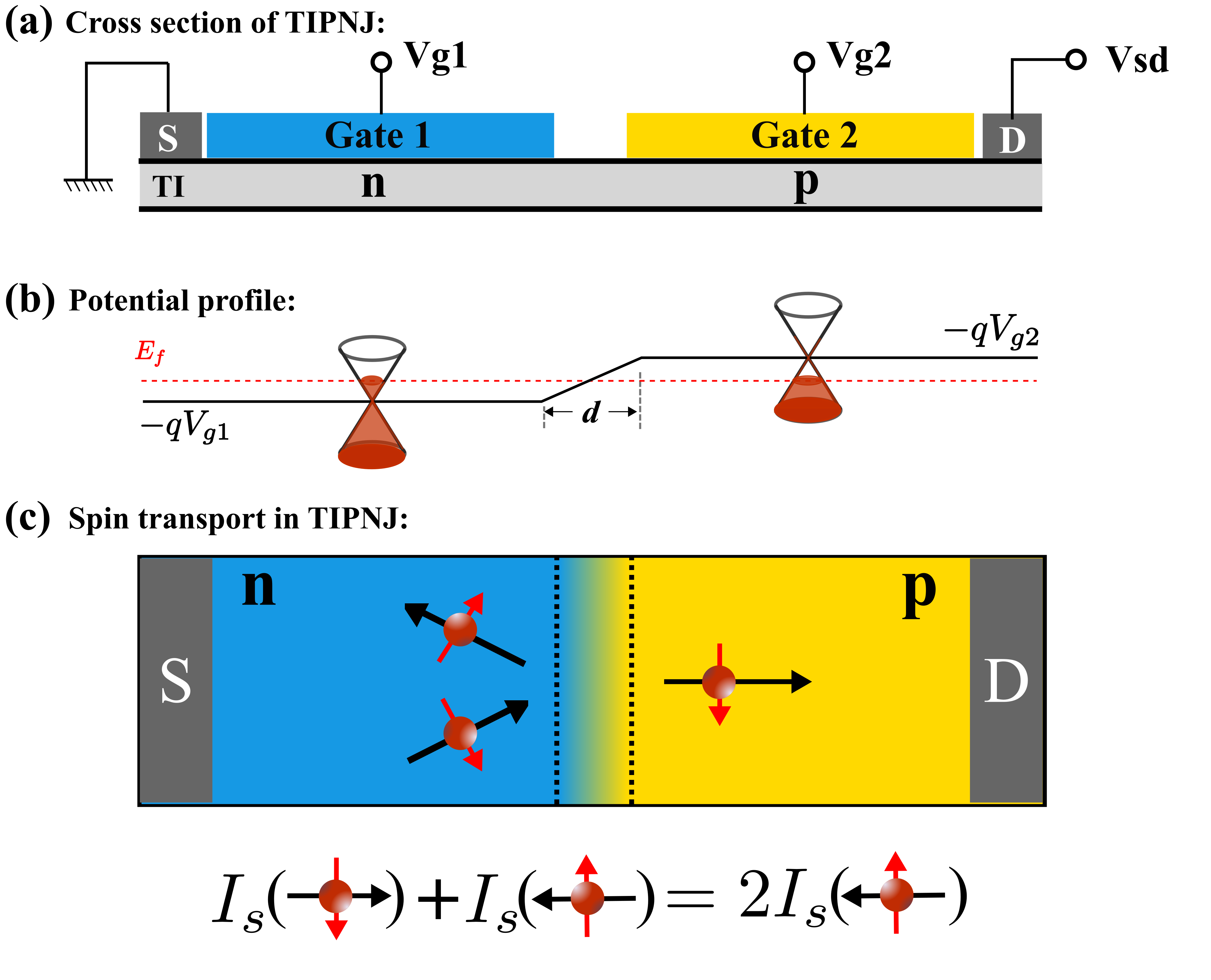}
\caption{Spin transport in topological insulator PN junction. (a) Side view of the setup for gate-controlled TI surface. (b) Electrostatic potential profile of the TI surface in a PN junction setup. (c) Schematic of spin transport in TI pn junction.}
\label{fig:TIPNJ}
\end{figure}
One possible way to achieve spin-to-charge gain is through 3D topological insulator surface states. For instance, we can use a gate controlled PN junction on a TI surface to control the spin-to-charge conversion\cite{habib2015chiral}. Fig.\ref{fig:TIPNJ} illustrates the idea of spin amplification at a TI PN junction. The TI surface is electrostatically doped into an N region and P region by top gates. An NEGF simulation based on a parameterized TI surface Hamiltonian shows that the PN junction acts like a `collimator', where only electrons with small incident angle with matching spins are allowed to tunnel through the junction. This collimation follows the physics of Klein Tunneling of Dirac fermions, also observed for pseudospins at a graphene PN junction. On the source side, most electrons with large incident angle are reflected back with their spins flipped due to the spin-momentum locking\cite{qi2011topological} of the TI surface states. While the charge current is reduced due to the backscattering of electrons, the spin current is amplified because of the simultaneous flip of momentum and spin. 

\begin{figure}[ht]
\centering
\includegraphics[width=8cm]{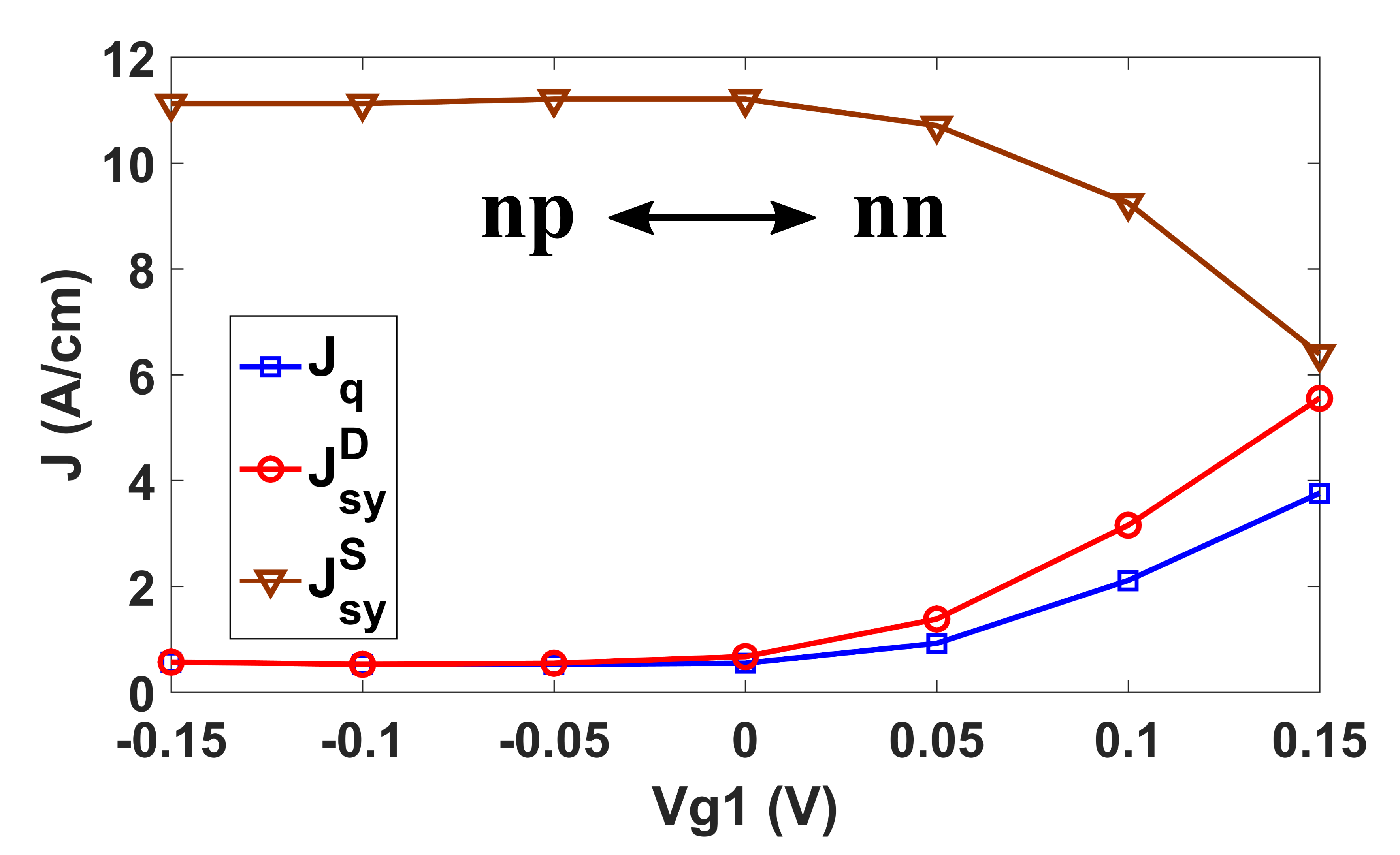}
\caption{Charge and spin current (source side and drain side) as a function of gate voltage $V_g1$ on the left side in Fig.\ref{fig:TIPNJ}. The gate voltage at the N side (right side) is fixed at $V_g2$ and the source-drain bias is at $V_{SD}=0.1\,\mathrm{V}$. The simulated TI surface has dimensions $200\,\mathrm{nm}\times 120\,\mathrm{nm}$ with a splitting $d=100\,\mathrm{nm}$ between gate 1 and gate 2. Other details can be found in ref.\cite{habib2015chiral}.}
\label{fig:TI_Curr}
\end{figure}

Fig. \ref{fig:TI_Curr} shows the NEGF simulation of the charge/spin transport on TI PN junction. The simulation result indicates a spin-to-charge gain that can be tuned by the gate voltage from $I_{sy}^{S}/I_q\sim 1.5$ to as high as $I_{sy}^{S}/I_q\sim 20$ (compared to $I_s/I_q\sim 2$ in usual GSHE through the geometrical gain). The large gate tunable spin to charge conversion is potentially useful for charge spin logic \cite{behin-aein_proposal_2010}. However, it is worth emphasizing that the transport focused exclusively on the surface conduction on TI. A more detailed model is needed to explore the spreading resistance of the surface currents into the bulk, to quantify any additional leakage of spin current. 

\section{Magnetization dynamics: stochastic vs deterministic switching}
\label{sec:reliability}
Once a spin polarized current enters the soft magnet, the spin torque tries to destabilize the magnetization, making it precess for small currents and eventually reversing the magnetization beyond a critical current. The dynamics is complicated and involves the current driven torque, the Gilbert damping that tries to reset the magnetization back towards its equilibrium value, and various real and pseudo-magnetic fields that arise from the potential that the spins swing in - from external fields, shape  and magnetocrystalline anisotropy, as well as thermal noise. The effect of thermal noise is particularly important. When the applied current is slightly below the critical current $I\leq I_c$, thermal excitation dominates the switching. At equilibrium (zero current), random kicks from the thermal noise create a probability distribution of the initial magnetic moment. These thermal fluctuations help the switching process by knocking the spins out of `stagnation', when the two magnets are momentarily locked into a state with precisely parallel or antiparallel magnetization (i.e., zero torque). Subsequently, thermal fluctuations hinder the smooth ballistic flow of spins under spin torque by creating occasional thermally induced backflips. 

The evolution of magnetization $\vec{m}(\theta,\phi,t)$ under spin transfer torque can be described by the Landau-Lifshitz-Gilbert-Slonczewski (LLGS) equation:
\begin{equation}
\begin{split}
 \frac{\partial \mathbf{m}}{\partial t}&=-\frac{1}{1+\alpha^2}\left[\mu_0\gamma\mathbf{m}\times\mathbf{H}_\mathrm{eff}+\alpha\mu_0\gamma\mathbf{m}\times\left(\mathbf{m}\times\mathbf{H}_\mathrm{eff}\right)\right.\\
  &\left.\qquad\qquad\qquad+\tau^\mathrm{stt}_{||}+\tau^\mathrm{stt}_{\perp}\right]\\
  \end{split}
    \label{eq:LLGS}
\end{equation}
where $\tau^\mathrm{stt}_{||}(\tau^\mathrm{stt}_{\perp})$ is the in-plane (out-of-plane) spin transfer torque, $\alpha$ is the magnetic damping coefficient, $\mu_0$ is the permeability constant and $\gamma$ is the gyromagnetic ratio. In PMA systems, the out-of-plane torque $\tau^\mathrm{stt}_{\perp}$ can be ignored and the in-plane torque can be approximated as $\tau^\mathrm{stt}_{||}=\mu_BI\eta\mathbf{m}\times\left(\mathbf{m}\times\mathbf{I}_s\right)/q\Omega M_s$ where $I$ is the charge current and $I_s$ is the unit vector along the direction of the polarized spins, $\Omega$ is the volume of the magnet, $M_s$ is the saturation magnetization and $q$ is the electron charge. $H_\mathrm{eff}$ includes the external magnetic field, the anisotropy field, and the demagnetization.  To describe the effect of  thermal noise, a noise field $H_\mathrm{th}=\sqrt{2\alpha k_BT/(\mu_0\gamma\Omega M_s)}\mathbf{G}$ is also included in the effective field $H_\mathrm{eff}$, where $\mathbf{G}$ is a three-dimensional Gaussian white noise uncorrelated in time and space. From the LLGS equation one can work out the zero temperature critical current $I_c=2q\alpha(2k_BT\Delta)/(\eta\hbar)$ (for PMA) where $\Delta=K_u\Omega/(2k_BT)$ is the thermal stability factor and $K_u$ is the effective magnetoanisotropy that includes the intrinsic anisotropy and the demagnetization along the perpendicular direction. Fig. \ref{fig:SLLG_vs_FPE}(left) shows the initial stagnation of the spins until thermal fluctuations dislodge them, and the subsequent evolution of the spins under the action of the current torque along with a superposed thermal jitter. It is worth mentioning that describing the magnet dynamics with a single LLGS equation assumes coherent switching, which works for magnet with size smaller than its exchange length. For a large magnet, the magnetization switching is mostly incoherent. Describing such complicated processes usually requires full micromagnetics simulations. Section 6.3 summarizes some efforts to approximate the incoherent switching effect within macrospin framework.

\subsection{Fokker-Planck equation}
\begin{figure}[ht]
\centering
\includegraphics[width=8cm]{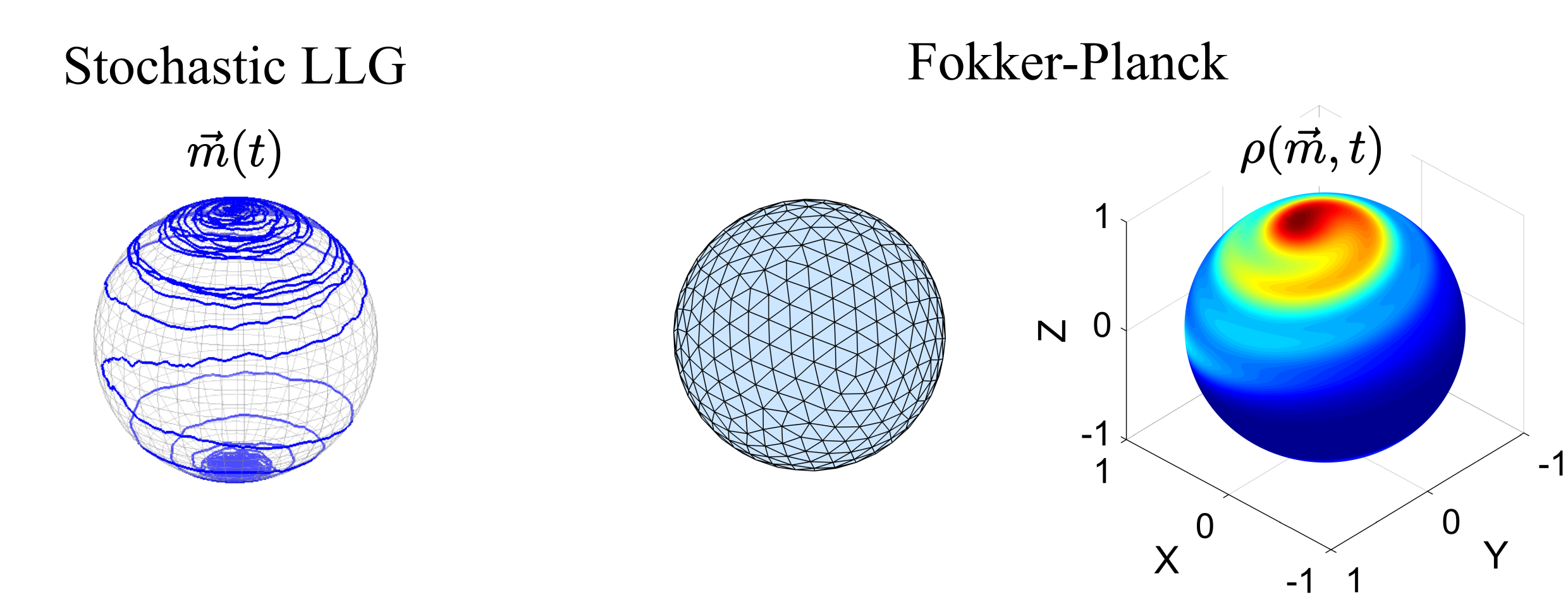}
\caption{Left. One shot stochastic LLG simulation of $\mathbf{m}(t)$. Middle. Uniform triangular meshes on a spherical surface. Right. A snapshot of the probability density $\rho(\mathbf{m},t)$ from the 2D Fokker-Planck simulation.}
\label{fig:SLLG_vs_FPE}
\end{figure}

An alternative way to describe the stochastic switching behavior is through the Fokker-Planck equation (FPE). Instead of keeping track of the noisy trajectory of the magnetic moment $\mathbf{m}(\theta,\phi,t)$ under thermal jitter, the FPE solves directly for the probability distribution of the magnetic moment $\rho(\mathbf{m},t)$. Under the macrospin assumption, the amplitude of the total magnetic moment is conserved. $\rho(\mathbf{m},t)$ can then be solved on a 2D spherical surface (Fig.\ref{fig:SLLG_vs_FPE} right). Brown\cite{brown1963thermal} first applied the FPE method to study the thermal excitation in single domain magnets. 

\begin{figure}[ht]
\centering
\includegraphics[width=8cm]{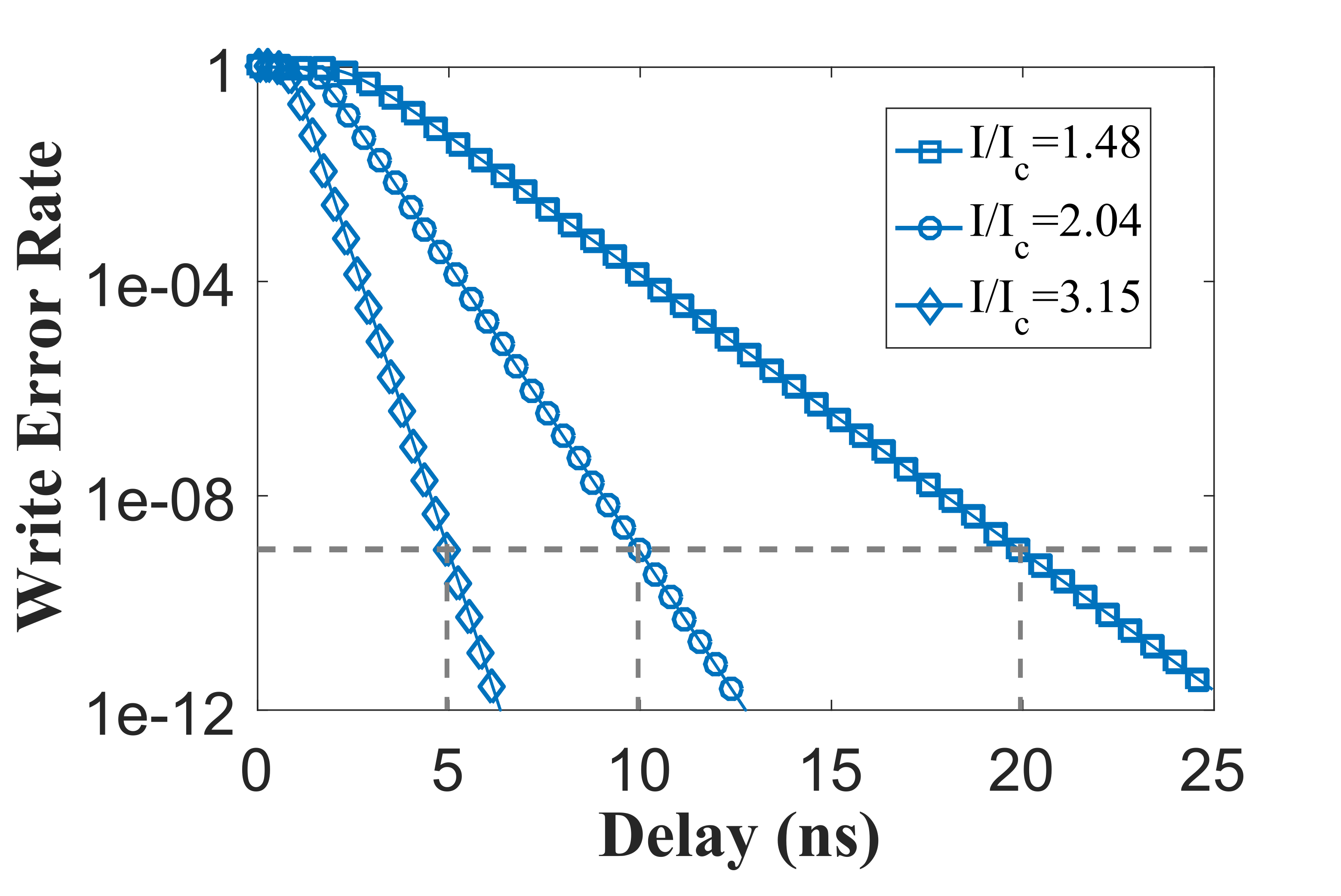}
\caption{Fokker-Planck simulation of the Write Error Rate (WER) as a function of the switching delay.  Three different applied currents are plotted, which hit the target WER=1E-9 at $5\,\mathrm{ns},\;10\,\mathrm{ns},\;20\,\mathrm{ns}$ respectively. The parameters used $M_s=1257\,\mathrm{emu/cc},\;H_k=3.34\,\mathrm{kOe},\;\alpha=0.02,\;\Delta=60$ are similar to those reported in \cite{ikeda2010perpendicular}. }
\label{fig:WER-t}
\end{figure}

It is straightforward to extend FPE to include spin transfer torque. We can derive the Fokker-Planck equation starting from the stochastic LLGS equation, assuming an uncorrelated Gaussian white noise for the thermal scattering processes \cite{ghosh2015nanoelectronics}. The Fokker-Planck equation can be expressed in a general form:
\begin{equation}
\begin{split}
\frac{\partial \rho}{\partial t}&=-\mathbf{\nabla}\cdot\left(\mathbf{L}\rho\right)+D\mathbf{\nabla}^2\rho\\
D&=\frac{\alpha\gamma k_BT}{(1+\alpha^2)\mu_0M_s\Omega}
\end{split}
\label{eq:2D_FP}
\end{equation}
where $\mathbf{L}$ is the sum of all the deterministic torques from the right side of Eq.\ref{eq:LLGS}. $D$ is the effective diffusion coefficient that captures the effect of thermal noise. Fig. \ref{fig:SLLG_vs_FPE}(Right) shows the evolution of the entire probability density function over time.  
Fig.\ref{fig:WER-t} shows a typical plot for the write error rate (WER) as a function of the switching delay for varying degrees of overdrive current.

\begin{figure}[ht]
\centering
\includegraphics[width=8cm]{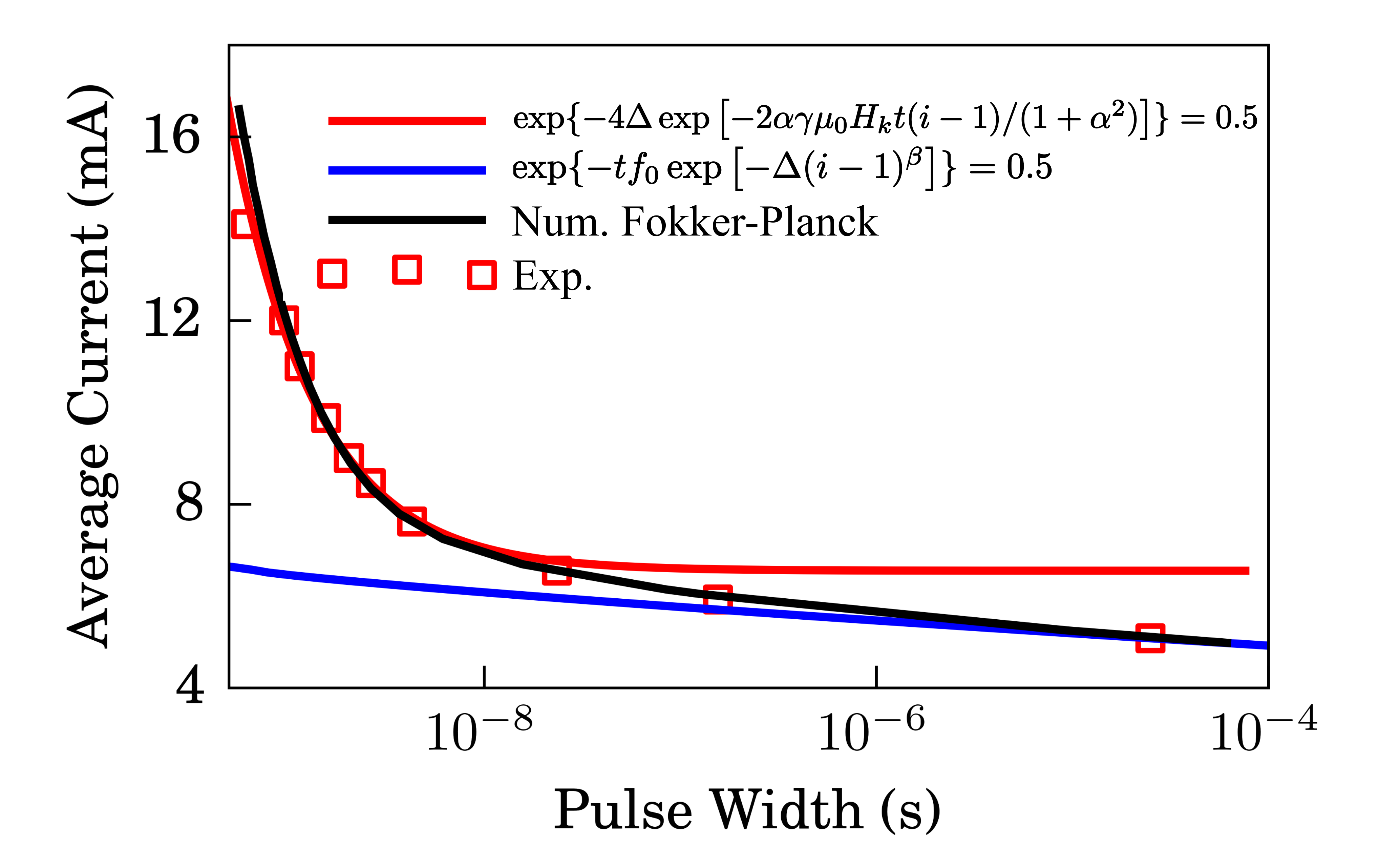}
\caption{Average switching current as a function of the switching delay. The theoretical fittings are calculated at $P_{sw}=\mathrm{WER}=0.5$. In the large current regime the Sun's model is used while in the small current regime the Arrhenius model is used. $i=I/I_c$ is the scaled current. $f_0$ is an empirical frequency and $\beta$ is a geometrical parameter that is different for the in-plane and the perpendicular systems. The experimental data is extracted from a $100\,\mathrm{nm}\times100\,\mathrm{nm}$ perpendicular spin-valve from ref.\cite{liu2014dynamics}. The fitting details can be found in ref. \cite{xie2017fokker}.}
\label{fig:I-t}
\end{figure}

The FPE and stochastic LLGS methods are formally equivalent, for magnets driven by uncorrelated white noise. However, there are two practical advantages of the FPE approach: first, the FPE is a deterministic equation that can be solved more efficiently than the stochastic LLGS equation, especially for statistical quantities such as averages and standard deviations (rare events can also be captured by the tail of $\rho$ but need a hyperfine grid). The one-time solution to Eq.\ref{eq:2D_FP} gives the statistics of the ensemble of most likely switching events.  Second, it is possible to obtain analytical solutions to the FPE under certain simplifying conditions. For example, the Fokker-Planck equation reduces to a 1-D differential equation in cylindrically symmetric systems such as the PMA MTJs. The analytical solutions can then be worked out when the applied current is in either the sub-threshold ($I\ll I_c$) or the super-threshold limit ($I\gg I_c$) \cite{butler2012switching}. In a more general case, the FPE can be solved numerically on a spherical surface using the finite element method and can be shown to bridge these limits seamlessly \cite{xie2017fokker}. Fig.\ref{fig:I-t} shows the comparison between the fittings from the numerical FPE and  two other widely used analytical models in the literature.

It is instructive to look at the relation between the WER and the total charge flowing through the magnet, which can be calculated approximately from the Sun's equation shown in the inset of Fig.~\ref{fig:I-t} and Eq.~\ref{eq:PMA_Switch_current}. From the write-error rate WER, we can get the following equation 
\begin{equation}
WER \propto e^{\displaystyle - 2(Q-Q_C)/Q_C}, ~~~ Q_C = \displaystyle\frac{qM_S\Omega}{\mu_B}\Biggl(\frac{1+\alpha^2}{\eta}\Biggr)
\end{equation}
which means that the efficiency of switching ultimately depends on the total accumulated charge
$Q$, and we can ramp up the accuracy by overdriving with charge exceeding the minimum critical
charge $Q_C$ to destabilize the spins towards flipping. The critical charge is obtained from simple
angular momentum conservation, trading off the spin $\mu_BQ_C/q$ with that from the magnets
$M_S\Omega$, while accounting for partial polarization of the tunneling electrons (making them less efficient by a factor $\eta$), and the damping correction $1+\alpha^2$ implying that part of the 
injected spin leaks out into the environment. For a magnet of size $100$nm $\times 20$nm with about
100,000 spins, we need about 10$^6$ electrons to provide the critical charge $Q_C$. For a $10$ ns
switching time, this already requires a current density of $\sim 1$ MA/cm$^2$.

\subsection{Examples: `Tilted' magnetization for fast switching}
As discussed earlier, one obstacle to fast, reliable spin torque switching in conventional magnetic tunnel junctions is the stagnation of the magnetization at points of zero torque. This initial incubation phase of STT plays a dominant role in determining the switching current, speed, and dynamic write error rate (WER). Several solutions have been proposed to overcome the stagnation point to achieve fast and reliable switching. One way to overcome this issue is a non-collinear alignment between the magnetic moments of the fixed layer and the free layer. 

\subsubsection*{Example 1 - orthogonal torque:} 
\begin{figure}[ht]
\centering
\includegraphics[width=8cm]{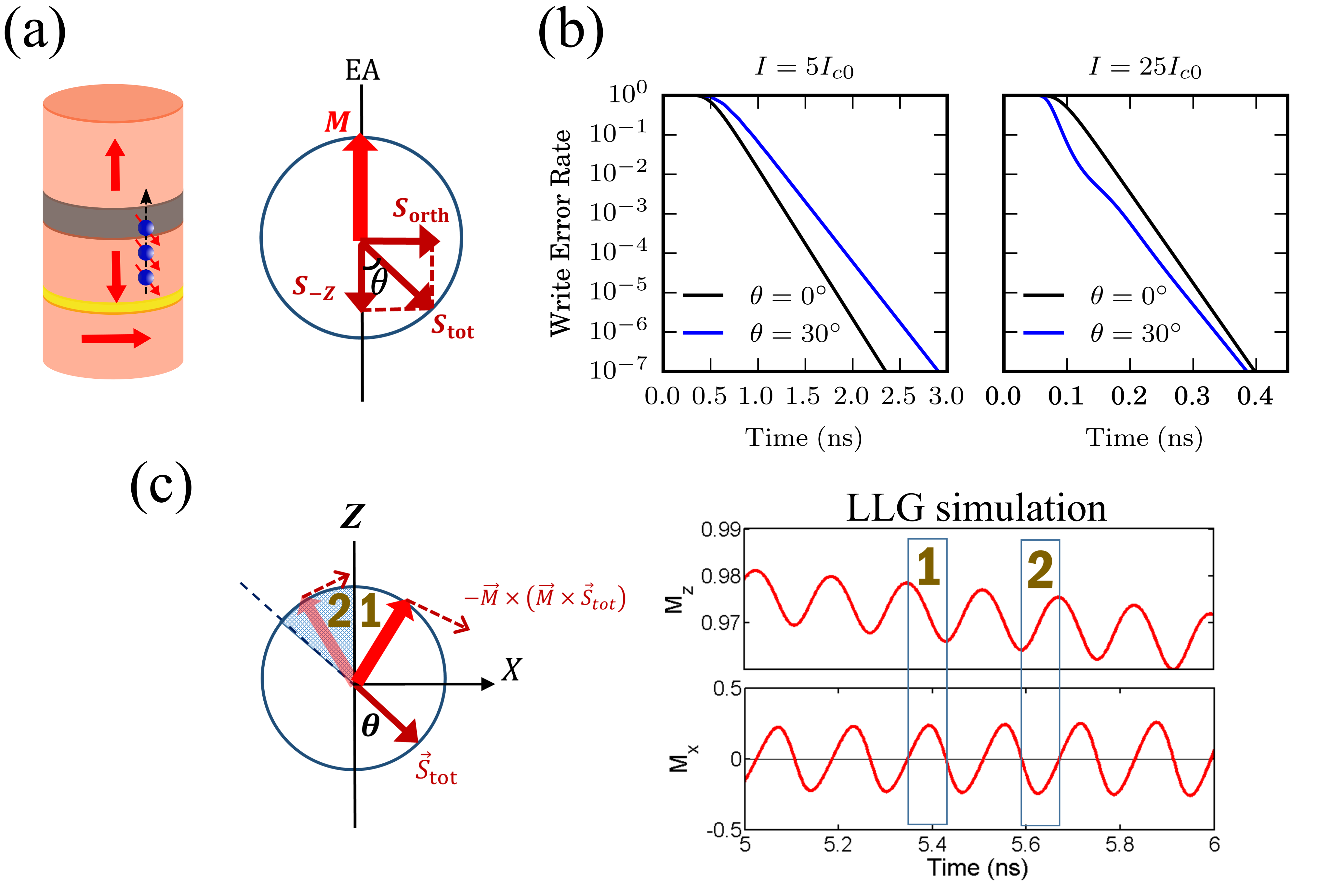}
\caption{(a) Schematic of creating non-collinear spin injection through an extra magnetic layer with an in-plane magnetic moment. By vector addition, we can think of the total effect as a non-collinear torque (with angle $\theta$ to the easy axis) injected
into the free magnet. (b) Fokker-Planck simulations comparing collinear and non-collinear spin injections for different applied current. (c) Schematic illustration of different regimes. The direction of the spin transfer torques $-\mathbf{M}\times(\mathbf{M}\times\mathbf{S}_\mathrm{tot})$ are indicated in different regimes in dashed arrows. LLG simulations of the $m_x(t),\,m_z(t)$ at the initial stage of switching. $m_z\rightarrow -1$ implies the STT helps to switch while $m_z\rightarrow +1$ implies the STT is against switching.}
\label{fig:orth}
\end{figure}

One possible way to nudge the magnetization out of its stagnation point is to use an orthogonal spin polarizer, which creates a non-collinear spin injection into the free layer as shown in Fig.\ref{fig:orth}(a). This method was used to achieve fast switching in in-plane MTJs \cite{liu2010ultrafast}. However, for PMAs there seems to be a trade-off between speed and current density. Fig.\ref{fig:orth}(b) shows a FP simulation of WER as a function of switching delay for collinear $\theta=0^\circ$ and non-collinear $\theta=30^\circ$ spin injections. When the injected current is small, the non-collinear spin injection has a worse WER than the collinear spin injection. The non-collinear spin injection only performs better when the applied current is much larger than the critical current, which is in the dissipative and ultrafast switching regime. 

We can understand the trade-off from the schematic figure and LLG simulation in Fig.\ref{fig:orth}(c): at small applied current, the magnetization switching in PMAs is precessional. At a small angle in the initial stage, a non-collinear torque helps to switch (regime 1) in the first half of the precession cycle, but it brings the magnetic moment right back towards its initial state in the second half (regime 2). This can be avoided only when the current is large enough to allow the magnetic moment to cross the equator in regime 1 itself. Such a large current is hard to achieve in an MTJ and creates undesirable dissipation. Other possible ways to generate an orthogonal torque could include a combination of GSHE and STT where the orthogonal torque can be generated from the GSHE without passing a large current through the MTJ. 

\subsubsection*{Example 2 - Easy-cone magnets:}
\begin{figure}[ht]
\centering
\includegraphics[width=8cm]{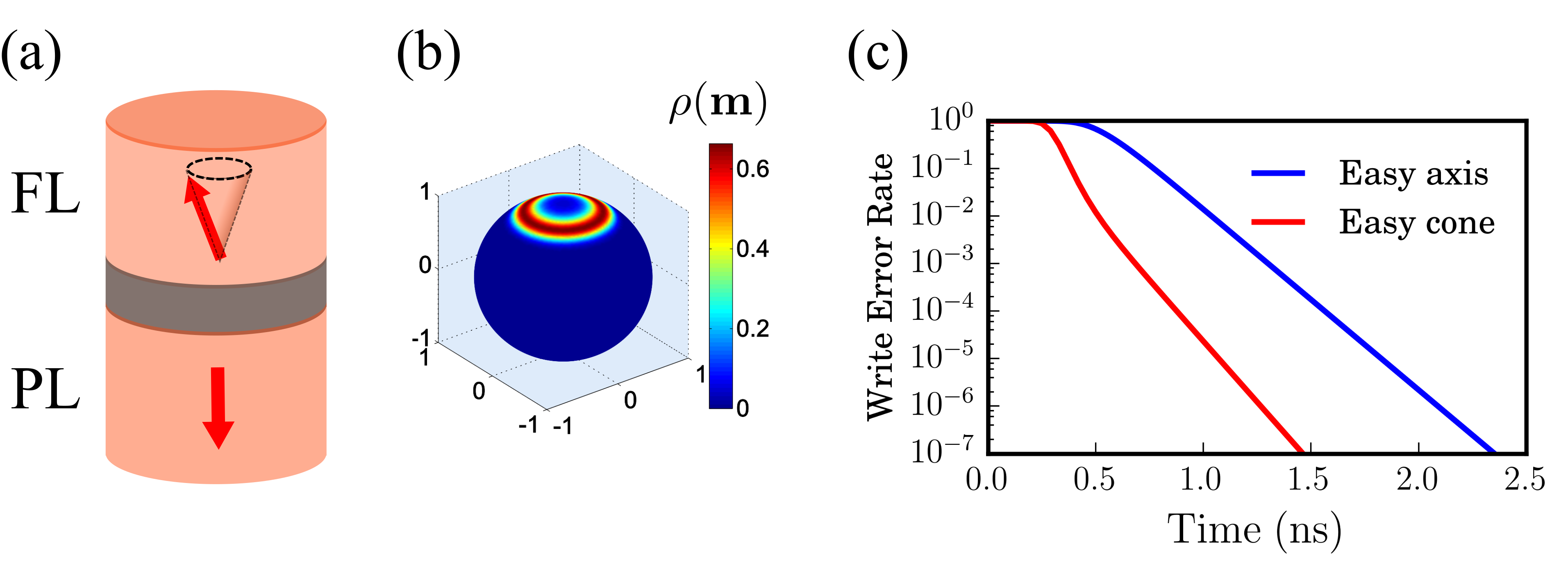}
\caption{(a). Schematic of an MTJ with a free layer that has the easy-cone magnetoansiotropy. (b). The equilibrium probability distribution of the magnetic moment in an easy-cone magnetic structure. $\Delta=43k_BT,\;K_4=-1.25K_\mathrm{eff}$ are used for the easy cone case. (c). Comparing the WER as a function of the switching time for easy-cone and easy-axis magnetic structure. The easy-axis case is set to the same energy barrier $\Delta=43k_BT$. The applied current is set $I=5I_c$ in both cases, where $I_c$ is the critical current for the easy-axis device.}
\label{fig:WER_easy_cone}
\end{figure}

The second example of non-collinear alignment is a system with conical magnetoanisotropy. In these systems, the energy profile can be expressed as $E=-K_\mathrm{eff}\cos^2\theta-0.5K_4\cos^4\theta$, where $K_\mathrm{eff}$ is the sum of all second order anisotropy terms including the usual interfacial perpendicular anisotropy and the demagnetization, while $K_4$ is a higher order anisotropy term. It has been shown that when the free layer $\mathrm{CoFeB}$ is within a certain thickness range, the energy minimum appears along the surface of a cone rather than along the axis \cite{shaw2015perpendicular}. Fig. \ref{fig:WER_easy_cone} shows the FPE simulation of the WER for an easy-cone magnet compared to an easy-axis magnet with the same thermal stability factor $\Delta$. As expected, a `tilted' initial magnetic moment helps reduce the switching time and error.
We see a quick initial evolution since the stagnation point of zero torque has been shifted. Subsequently we reach the shifted stagnation point, and after that the evolution slows down, but since the Boltzmann distribution for the easy cone is not along the stagnation point, we still see a faster slope of the WER.

\subsection{Non-coherent switching}
So far we have assumed that the magnet switches in a coherent way so that all the electron spins are bundled together as one `giant spin'. This approximation holds reasonably well for small sized single domain magnets, but there seems to be evidence that the transient response of the spins is not coherent, especially for intermediate sized magnets. A full spatially resolved micromagnetic simulation is needed to describe the complicated magnetization dynamics in these systems.  Studies have suggested various possible mechanisms for transient incoherence, such as sub-volume excitation \cite{sun2011effect}, edge nucleation \cite{finocchio2007micromagnetic} and global magnetostatic instability \cite{munira2015calculation}. The general idea behind incoherent switching is that a medium sized magnet is excited by the STT through multiple precessional modes with different frequency and spatial patterns. Among those modes, the simplest is the coherent uniform precession. However, to switch the magnet in a coherent mode requires the magnetic moment to overcome the energy barrier $E_b=K_u\Omega$ where $\Omega$ is the whole volume of the magnet. Micromagnetic simulations suggest several switching mechanisms that only need to overcome a lower barrier height. For example, ref.\cite{munira2015calculation} suggests a global magnetostatic instability in STT switching in PMAs, where a non-coherent precessional mode starts to destabilize and diverge when the precession amplitude becomes large. Once destabilized, the entire magnet will eventually flip. Therefore, instead of torquing the entire magnetic moment across the equator ($\theta=\pi/2$), one just needs to excite the magnetization precession to pass the destabilizing angle,  which translates into a smaller barrier height. In the thermal excitation regime with STT, which is the regime where non-coherent flow becomes dominant, the anisotropy energy modified by the spin torque at $\theta$ is $E(\theta)=K_u\Omega(2I\cos\theta/I_c-\cos^2\theta)$ \cite{munira2015calculation}.  For coherent switching, the energy barrier is given by $E_b=E(\theta_\mathrm{max})-E(0)=K_u\Omega(1-I/I_c)^2$ where $\theta_\mathrm{max}$ is the angle where the anisotropy energy maximizes (no longer at $\theta=\pi/2$ because of STT but at   $\theta_{max}=\cos^{-1}(I/I_c)$, obtained by globally maximizing $E(\theta)$. But the instability implies a lower energy barrier height $E_b'=E(\theta_{sw})-E(0)$ where $\theta_{sw}$ is the instability angle and $E(\theta_{sw})\le E(\theta_\mathrm{max})$. One can then use the modified energy barrier $E_b'$ in a macrospin model to evaluate the switching rate and switching error approximately.

\section{Emerging Write Mechanisms For Nano-Magnets}

\label{sec:emerging_device}

\begin{figure}[ht!]
\centering
\includegraphics[width=8cm]{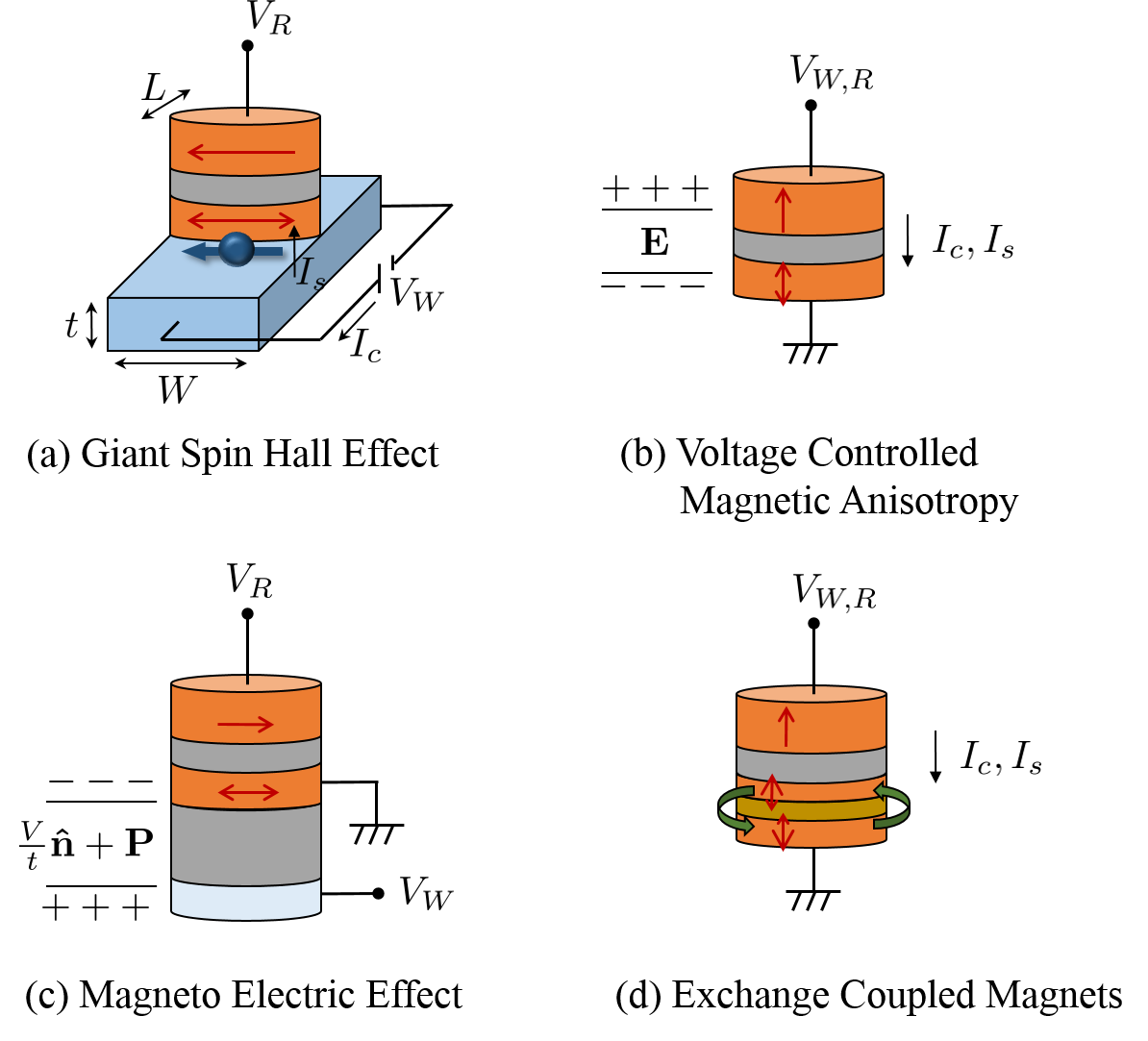}
\caption{(a) Giant Spin-Hall Effect (GSHE) based switching of nano-magnet allows a 3-terminal design of MRAM cells with separate paths for read and write operations. Due to the aspect ratios of the GSHE film and the nano-magnet, it is possible to get a net gain $\eta \gg 1$ in the write operation, compared to pure STT based switching. (b) Voltage Controlled Magnetic Anisotropy (VCMA) effect can assist in lowering the minimum spin current needed to write to MRAM cells by dynamically lowering the anisotropy field strength through the electric field $\mathbf{E}$ developed across the junction. (c) Magneto-Electric (ME) effect allows for voltage based write operation in an MRAM cell by generating a controllable magnetic field coupled with the controllable polarization of the multiferroic "gate" at the bottom of the MTJ. (d) Using two complementary Exchange-Coupled Magnets (ECM) as the free layer of the MRAM cells allows for total thermal stability equivalent to the sum of the thermal stabilities of both the magnets, but the spin-torque current needed is equivalent to a magnet with net difference of the anisotropy of the two coupled magnets.}
\label{fig:emerging}
\end{figure}

The last half a decade or so has seen an explosion of materials and novel switching mechanisms for manipulating nanomagnets. While "plain old" STT has already seen commercialization in second generation Magnetic RAMs as STT-MRAMs, these new mechanisms are still experimental and in various stage of commercial development. Some of the experimentally demonstrated materials and phenomena are (Figs.
\ref{fig:emerging},~\ref{fig:pma_ima})
\begin{enumerate}
\item Nano-magnets with interfacial Perpendicular Magnetic Anisotropy (PMA)
\item Giant Spin-Hall Effect (GSHE)
\item Voltage Control of Magnetic Anisotropy (VCMA)
\item Magneto-Electric Switching (ME)
\item Exchange-Coupled Magnets (ECM)
\end{enumerate}

This section briefly describes the physics behind these switches, their intrinsic benefits, and challenges behind each of these materials and/or phenomena. Other mechanisms not discussed here include domain-wall and skyr\-mion-based devices, piezoelectric/strain assisted switching, magnonic (spin-waves) and thermomagnonic switching.

\subsection{Magnets with Perpendicular Magnetic Anisotropy}

\begin{figure}[ht!]
\centering
\includegraphics[width=8cm]{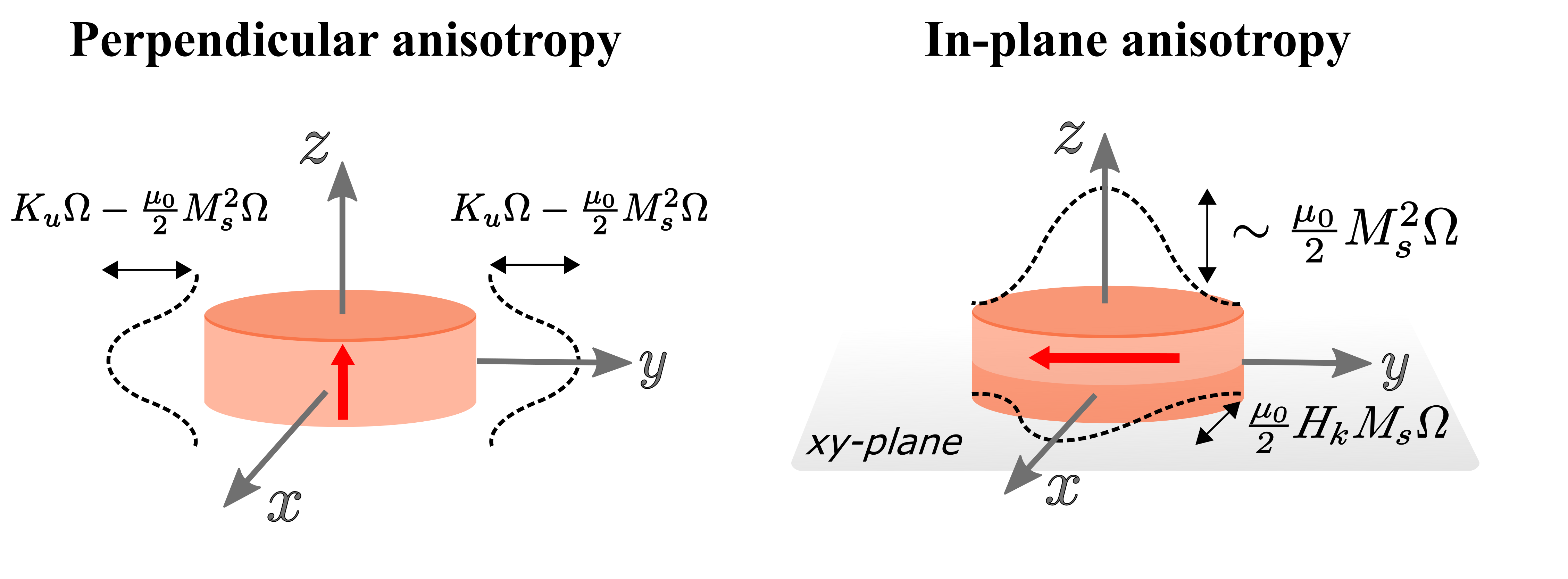}
\caption{Left. Anisotropy energy barrier landscape in a PMA magnet. The easy axis is the $z$ axis. The switching process has to overcome the barrier in the plane of the magnet and is given by a single effective anisotropy field which is the difference of the "bare" anisotropy $K_u \Omega$ and the demagnetization $\mu_0M_s^2\Omega/2$. Right. The energy landscape of an In-Plane-Anisotropy (IMA) magnet. The easy axis is the $y$ axis. The switching process negotiates two different barriers in-plane $\mu_0H_kM_s\Omega/2$ and out-of-plane directions $\mu_0M_s^2\Omega/2$.}
\label{fig:pma_ima}
\end{figure}

\textit{Physics:} A magnet's anisotropy can be caused by a large number of factors \cite{johnson1996magnetic}, one of the important  being its shape. Magnetic films of elliptical/rectangular profile and thickness of a few nm primarily have their easy axis in the direction of the major axis of the ellipse or long side of the rectangle, with the film plane being the easy plane of the magnet. In each of these descriptions `easy' denotes the direction in which the potential energy is minimum for the magnet. This potential energy profile prohibits the magnetization to wander out to the `out-of-the-plane' axis thermodynamically and is typically included in magnetodynamics calculations as an effective demagnetization field. However, films with circular profiles and extreme thinness ($t<2\ nm$) primarily have a high preference for anisotropy in the out-of-the-plane or perpendicular to the plane direction due to surface interactions, especially at the $\rm CoFeB|MgO$ interface \cite{ikeda2010perpendicular}. This high anisotropy can overcome the demagnetization field and make the perpendicular direction the thermodynamically stable point for magnetization (Fig.~\ref{fig:pma_ima}).

PMA's have gained a lot of attention recently due to their scalability. Since the magnets are circular in profile and their high anisotropy originates from film thickness, it is lithographically possible to create magnets with very large anisotropy with a very short footprint with high stability, and is now routinely adopted in the magnetic storage industry. The uptake of PMA magnets by the STT-MRAM industry is on the lines of the storage industry trends.

\textit{Benefits to Switching:} It has been shown that, assuming a mono-domain texture, a magnet's minimum switching current is given by \cite{Sun2000}:

\begin{equation}
I_{min.} = \frac{2q\mu_0}{\hbar}\frac{\alpha}{\eta}M_s \Omega (H_k + \frac{M_s}{2}) 
\end{equation}

\noindent where once again, $\alpha$ is the Gilbert damping parameter, $\eta$ is the effective spin-current polarization, $M_s$ is the saturation magnetization of the magnetic material, $H_k$ is the anisotropy field strength, and $\Omega$ is the magnet's volume ($M_s\Omega$ is the total magnetic moment).

In PMA magnets, due to the lack of the out-of-the-plane demagnetization field (see fig. \ref{fig:pma_ima} for a visualization), the minimum current is:

\begin{equation}
I_{min.} = \frac{2q\mu_0}{\hbar}\frac{\alpha}{\eta}M_s \Omega H_k
\label{eq:PMA_Switch_current}
\end{equation}

The difference can be substantial for magnets with large saturation magnetization values, such as $\rm CoFeB$. Reduction of the drive current without any reduction in the stability of the magnets is beneficial in reducing the total Joule dissipation $I^2R$ of the drive circuitry. 

\textit{Challenges:}
PMA magnets are rapidly becoming a platform of choice for MRAM industry due to their scalability. However, challenges with high write error and stalling of the initial angle remain for the PMA magnets driven with Spin-Transfer Torque (STT). Another critical challenge to the PMA magnet based MTJs is scaling up write current density with scaling down of bit size, which can be understood in the following way.

For a monodomain magnet, the expected thermal stability time $\tau$ is:

\begin{equation}
\tau = \tau_0 e^{\Delta/kT}
\label{eq:tau_magnet}
\end{equation}

\noindent where $\tau_0\sim 0.1-1\ ns$ usually. Using  $\Delta = \mu_0 M_s \Omega H_k/2 \sim 60 kT$ gives us a decade of state retention which is a typical design target for memory applications. Even if the magnet area scales down the write current is fixed for a fixed $\Delta$ as evident from eq. \ref{eq:PMA_Switch_current}. Therefore as the volume $\Omega$ is reduced by scaling down the diameter $d$, the current density $\mathbf{J}$ increases $\propto 1/d^2$. Considering that the resistance of MTJ junctions also increase with scaling down the cross section area, scaling poses reliability and longevity challenges of electro-migration and dielectric breakdown of the MgO oxide layer.

\subsection{Giant Spin-Hall Effect}

\textit{Physics:} It is well known that materials with  Spin-Orbit Coupling have internal magnetic fields \cite{dyakonov1971current} that can give rise to intrinsic spin-polarization and spin-Hall effect \cite{sih2005spatial}. It has also been demonstrated that the spin accumulation at the interface of a heavy metal (Pt, W, $\rm \beta-Ta$) and a magnet (CoFeB) can reliably switch the magnetization of the magnet \cite{liu2011spin}. This has been explained either through the mechanism of Dzyaloshinskii-Moria (DM) interaction \cite{miron2010current}, or through the accumulation of spins at the interface, which can then act as a virtual "spin-battery" from which it is easy to inject spins \cite{pai2012spin} compared to semiconductor-metallic magnet interfaces due to better impedance matching in all metallic material system. 

\textit{Benefits:} The primary motivation for GSHE based switching is the spin-current gain that can be obtained through geometry. The GSHE is given as $\theta_{SHE} = |\mathbf{J_s}|/|\mathbf{J_c}|$ where $\mathbf{J_s},\mathbf{J_c}$ are the spin-current and charge-current densities in the transverse and the longitudinal directions respectively. Given that the charge current flows through (see fig. \ref{fig:emerging}a) the GSHE film interface through the cross-section $W\times t$ and spin current flows into the magnet through the cross-section $W\times L$, the current ratios, or in other words effective polarization (for $t\gg \lambda$, spin flip length in GSHE), is given as:

\begin{equation}
\eta_{SHE} = \frac{I_s}{I_c} = \frac{|\mathbf{J_s}|WL}{|\mathbf{J_c}|Wt} = \theta_{SHE}\frac{L}{t}
\end{equation}

\noindent which can be controlled by device design and it is possible to obtain a polarization $\eta \gg 1$ which can be viewed as a `gain' from an engineering perspective. This can be intuitively understood through constant depolarization-repolarization of itinerant electrons at the interface of the GSHE$|$Magnet films, due to the large internal field generated by large Rashba effect in the GSHE material. This constant depolarization-repolarization allows a single electron to impart multiple units of spin-torques to the magnet, as opposed to a spin injected from a fixed polarizing layer in MTJ or spin-valves.

\textit{Challenges:} GSHE materials however have a challenges that can limit their performance. Some of these are: 

\begin{enumerate}
\item Magnet scalability: as described above, the gain is due to the geometry of the GSHE$|$Magnet structure. This imposes a limit to the scalability of the magnets that can be switched without sacrificing the gain. Reducing the length $L$ of the magnet reduces the gain $\eta$ whereas the concomitant reduction of the width $W$ of the GSHE$|$Magnet film system increases the resistance of the whole structure, causing higher dissipation ($I^2 R$ losses). 

\item GSHE scalability: The GSHE film thickness cannot be scaled down indefinitely. For films with $t \sim \lambda$ (spin-flip length) the expression for gain is \cite{hong2016spin}:

\begin{equation}
\eta_{SHE} =  \theta_{SHE}\frac{L}{t}(1-sech(\frac{t}{\lambda}))
\label{eq:GSHE_gain}
\end{equation}

\noindent For $t\rightarrow 0, \eta_{SHE} \rightarrow 0$. This is explained by mixing of the opposite polarization of the spins on the top and the bottom of thin GSHE film surfaces (assuming the same θ for both) causing reduction in effective $\eta_{GSHE}$. In addition, it has been found empirically that in general materials with high $\theta_{SHE}$ tend to have high charge resistivity, which limits the materials that can be used for obtaining high GSHE without high $I^2 R$ losses.

\item Spin-current Shunting: The equation (\ref{eq:GSHE_gain}) for spin current injected into the magnet assumes that the magnet fully absorbs the entire spin current incident on it, i.e. it is a perfect ground for the spin current. However, magnets have finite resistance and for the spin-torque current, it is limited by the Sharvin resistance. Some amount of spins is reflected back from the GSHE$|$Magnet interface. The shunting conductance of the magnet is given by $g_0WL$ where $g_0$ is the interface/mixing conductivity. The GSHE spin shunting conductance is given by $\frac{\sigma WL}{t}tanh(2t/\lambda)$. These two shunting paths share the spin current generated by the GSHE, reducing the spin-injection efficieny into the magnet. 

\item Charge-current Shunting: The charge resistivity of typical GSHE materials (say $\beta$-Ta) is higher compared to the typical magnet material (say CoFeB). Therefore, a lot of the lateral charge current flowing in the GSHE gets shunted through the magnet rather than flowing through the GSHE, thereby reducing the effective amount of charge current available for spin-current generation.

\end{enumerate}

\subsection{Voltage-Controlled Magnetic Anisotropy}

\textit{Physics:} It has been demonstrated \cite{wang2012electric,shiota2012induction,kanai2014magnetization} that it is possible to modulate the surface anisotropy of ultra-thin PMA magnets by application of an electric field (fig. \ref{fig:emerging}b). This is caused primarily due to the reduction of the anisotropy field strength $H_k$. Therefore, with this effect, it is possible to dynamically lower the energy barrier of the magnet $\Delta$ and thereby reduce the write current necessary to switch the magnet (eq. \ref{eq:PMA_Switch_current}).

The VCMA effect is phenomenologically described by \cite{pertsev2013origin}:
\begin{equation}
k_{u,\perp}^{eff.} = k_{u,\perp}^0 -\eta_{A}\mathbf{E}
\end{equation}

The microscopic origins of this effect has been explained through accumulation/depletion of carriers at the interface of the Magnet$|$Oxide system. Spin-selective screening effects gives rise to a field dependent surface spin texture that changes the surface magnetic properties \cite{VCMAmicroscopicDFT}. The strength of the effect is commonly expressed in the units of $\frac{\mu J/cm^2}{V/nm}$. For a magnet of area of cross section $1\ nm^2$ and for an electric field strength of $1\ V/nm$ the reduction of the anisotropy energy is about $2.4\times 10^{-4} kT$ at room temperature. Values in excess of $\eta_A = 100\ \frac{\mu J/cm^2}{V/nm}$ have been reported \cite{wang2012electric}.

\textit{Benefits:} The primary benefit of the VCMA is in its ability to reduce the write current required to switch the magnet through the extra voltage bias on the magnet, which can almost be thought of as a gate voltage. Therefore, it is possible to dynamically control the magnet's volatility.

\textit{Challenges:} The following are the challenges that limit the VCMA effectiveness:

\begin{enumerate}
\item Magnet scalability: Being a surface effect, VCMA is proportional to the area of cross section of the interface and scales down as the magnet volume $\Omega$ is scaled.

\item Speed of switching: VCMA effect works by reducing the $H_k$ of the magnet, but leaves the $M_s\Omega$ product untouched. This means that the total charge that needs to be provided to switch the magnet ($=2qM_s\Omega/\mu_B$ \cite{camsari2016ultrafast}) does not change. Which means that the write current, though lower, needs to be provided for a longer time for full switching, therefore decreasing the speed of switching and potentially increasing the write error rate.
\end{enumerate}

\subsection{Magneto-Electric Effect}

\textit{Physics:} Multiferroics are a class of materials where multiple "ferro" phases coexist at the same time, and it is usually possible to manipulate one phase by controlling the other. Multiferroics are either single phase materials such as $\rm LaMnO_3, BiFeO_3$ etc. or they can be built through heterostructures (composite multiferroics) where the two ferro-phases interact through another intermediate ferro-phase.

It has been found that a large number of single phase multiferroics with ferroelectric order exhibit anti-ferromagnetism \cite{wang2010multiferroic}. However, in $\rm BiFeO_3$ (BFO), there is an uncompensated residual ferromagnetic order present (due to DM interaction) that is coupled with the ferroelectric order at room temperature \cite{ederer2005weak}. By controlling the ferroelectric order through an applied voltage bias it is possible to control the ferromagnetism of the ME material. This can then be used to generate an effective controllable magnetic field on an adjacent magnetic film mediated via exchange interaction \cite{borisov2005magnetoelectric}.

It has been experimentally demonstrated \cite{wu2010reversible,heron2014deterministic} that full reversible switching of a magnetic film grown over on top of a BFO is possible purely through voltage application on the BFO layer.

Basic phenomenological relationship of multiferroic phenomena is given by the magneto-electric tensor:
\begin{equation}
\alpha_{ME} = \mu_0\frac{d\mathbf{M}}{d\mathbf{E}}
\label{eq:ME_physics}
\end{equation}

Using the definition $\mathbf{B}=\mu_0 \mathbf{M}$ and $\mathbf{E}=({V}/{t})\mathbf{\hat n}$, we can relate the total magnetic field generated by the ME material to the total electric polarization. Note that this ignores the inbuilt polarization, which in principle will add dynamics of its own. The switching then is due to this generated magnetic field given by:
\begin{equation}
\mathbf{B}=\alpha_{ME}\frac{V}{t}\mathbf{\hat n}+ \alpha \mathbf{P}
\label{eq:ME_Switch}
\end{equation}

\noindent where $V$ is the applied voltage, $t$ is the thickness of the ME film, and $\mathbf{P}$ is the intrinsic polarization of the ME material. The ref \cite{heron2014deterministic} found a strong field dependence on $\alpha_{ME}$ and reported maximum value of $\alpha_{ME} \geq 10^{-7} s/m$.

\textit{Benefits:} ME based switching opens a pathway towards voltage based magnetization switching (e.g. see fig. \ref{fig:emerging}c as a possible structure), which is compatible with the current CMOS-based technology. The biggest advantage provided by the ME-based switching is the potential reduction of charge needed to switch the magnet. It can be shown from angular momentum considerations that the total amount of charge needed to switch a magnet is $2qM_S\Omega/\mu_B$ assuming $100\%$ polarization of spins \cite{camsari2016ultrafast}. Therefore, larger the $M_s\Omega$, higher the amount of charge needed, thereby having higher dissipation. However, the ME based switching, being in effect a field-like switching mechanism, needs to only provide sufficient amount of charge necessary to create the anisotropy field $H_k$ needed for switching the magnet which in principle can be much smaller and can be engineered by using a thin ME film \cite{ganguly2017evaluating}.

\textit{Challenges:} Some of the fundamental challenges with ME based switching are:

\begin{enumerate} 
\item Scalability of the ME material: This effect was demonstrated on BFO films that were substantially thick. The “gain” in this switching mechanism is due to the $\alpha_{ME}/t$ factor, which in theory can be very large for small $t$, enabling a small voltage to switch a high $H_k$ magnet. However, it is not understood if it is possible to scale down the ME material while keeping the ME effect ($\alpha_{ME}$) large. 

\item Scalability of the magnet: The principle of scaling of a magnet by reduction of the volume $\Omega$ needs to be balanced by increasing the $M_s$ or $H_k$ to maintain a $\Delta = 60\ kT$. For spin-torque-driven switching, it is beneficial to reduce the $M_s$ and increase $H_k$. However, for field based switching it is beneficial to increase $M_s$ and reduce $H_k$. However, while $H_k$ can be varied over a wide range of values by lithography and fabrication \cite{albrecht2013bit}, $M_s$ is determined by material composition and cannot be controlled through lithography. Therefore, scaling of ME based devices may require using complex exchange-coupled magnetic stacks employed by the storage industry for modern hard disk drives.

\item Switching dynamics poorly understood | potentially low speed and high write failure mechanism: The paper \cite{heron2014deterministic} deduced that the switching of the magnet was due to a complicated two-step rotation of the DMI-generated field, likely due to a complex interaction of the DMI and exchange interactions. Therefore, detailed micromagnetic studies and experiments are necessary to establish the physics of the effect. If the switching is a two-step phenomenon, it is likely to be a slow mechanism with high write failures.

\end{enumerate}





\subsection{Exchange Coupled Magnets}

\textit{Physics:} Instead of using a single large magnet with large anisotropy, it is possible to use a stack of exchange coupled magnets (fig. \ref{fig:emerging}d) where the total anisotropy adds up to provide a stable magnet. That is, the total magnetic barrier of the stack built of $n$ layers is given by:

\begin{equation}
\Delta = \left[\mu_0 H_k M_s \Omega/2\right]_{total} = \sum_{i}^{n} \left[\mu_0 H_k M_s \Omega/2\right]_{i}
\end{equation}
\noindent However, it was shown that if the magnet is made of two asymmetric layers with unequal but opposing/comp\-lementary magnetization directions, it is possible to switch the magnet with a much lower amount of charge, and is equivalent to switching a magnet with anisotropy that is the difference of the anisotropy of the two layers, given by \cite{camsari2016ultrafast}:

\begin{equation}
\left[\mu_0 H_k M_s \Omega\right]_{eff.} = \left[ \mu_0 H_k M_s \Omega\right]_{1} - \left[ \mu_0 H_k  M_s \Omega\right]_{2}
\end{equation}

\textit{Benefits:} The primary benefit that is derived from exchange coupled stacks is the reduction of the total charge needed to switch the overall stack's magnetization, without sacrificing the thermal stability.

\textit{Challenges:} While this technique is widely adopted in the storage industry in building magnetic bits of extremely small dimensions \cite{albrecht2013bit}, it has as yet not been adopted in the STT-MRAM industry due to lithography and fabrication challenges of building such complex stacks.

\section{Putting it all together: Integrated approach based on analytical models}
\label{sec:analy_modular}

\begin{figure*}[ht]
\centering
\includegraphics[width=18cm]{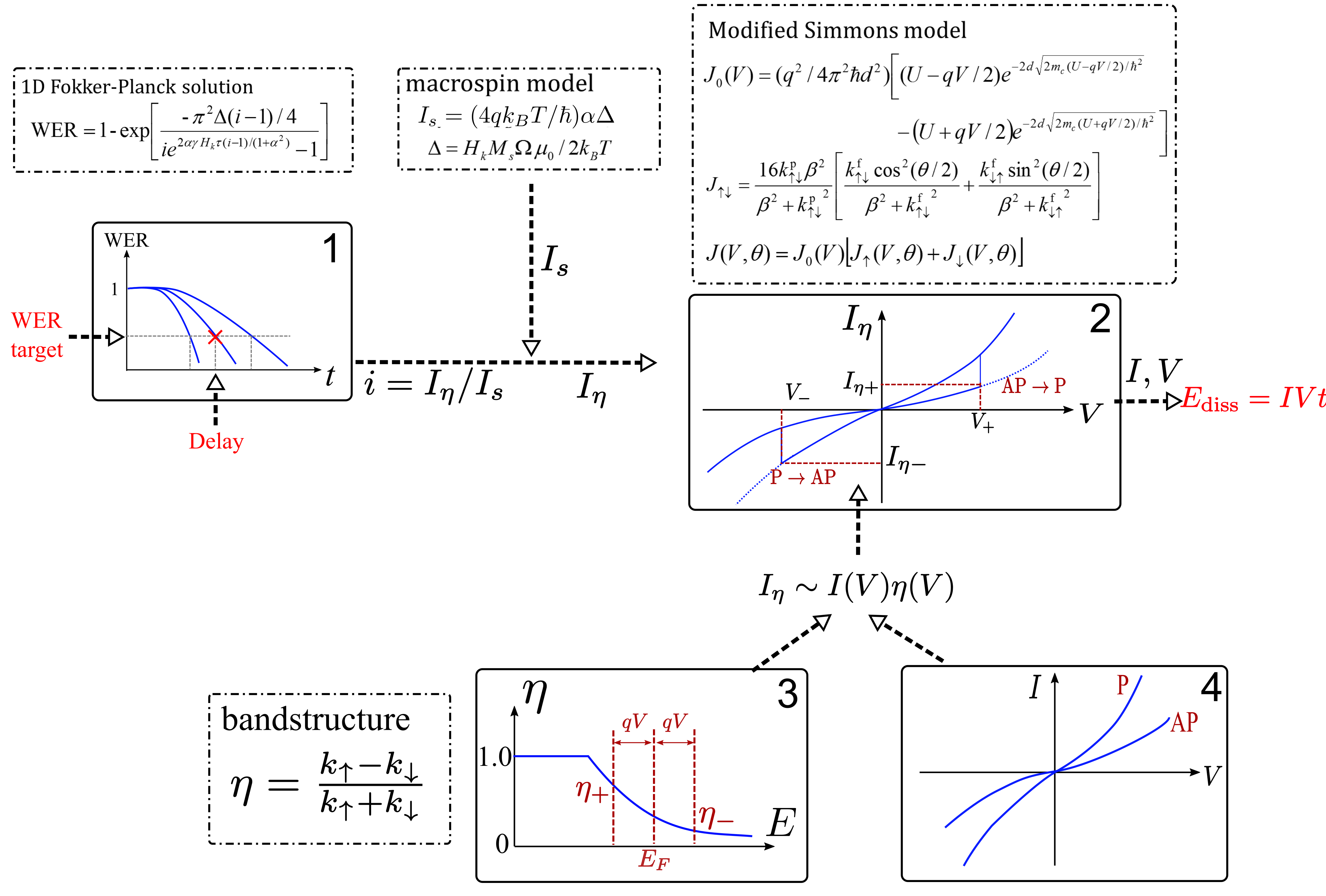}
\begin{enumerate}
\item 
\end{enumerate}
\caption{Energy-delay-error co-design diagram. The inputs are pulse width and target write error rate. The WER module (Block 1) calculates the effective current $i=I_\eta/I_s$ that meets the error target for a given pulse width, where $I_\eta$ is the polarized current and can be written as $I_\eta=I(V)\eta(V)$ in the simple case. $I_s$ is the critical current assuming $100\%$ polarization $I_s=I_c(\eta=1)$ from the macrospin model. To determine the voltage, we need to plot the polarized current $I_\eta$ as a function of voltage $V$ (Block 2). This plot comes from the combination of $\eta(V)$ (from the bandstructure module, Block 3) and $I(V)$ from transport module such as the modified Simmons equations (Block 4). Finally, the voltage and the charge current can be used to calculate the energy dissipation.}
\label{fig:diagram}
\end{figure*}

In this section, we use the magnetic tunnel junction as an example to show how to integrate different modules, each represented by an analytical compact model. Fig.\ref{fig:diagram} shows the diagram and the underlying equations for this integrated approach to study the energy-delay-error of STT switching\cite{munira2012quasi}. This process starts with a target write error rate and switching delay, which is usually application specific. The WER and $t$ are the input into an error model. Here we choose the analytical solution to the 1-D Fokker-Planck equation for WER in a perpendicular magnetic tunnel junction. In the 1D FPE model, WER is a function of switching delay $t$ and overdrive current $i=I_\eta/I_s$ as shown in the diagram. Instead of the usual way to express the overdrive current $i=I/I_c$, we use the polarized current $I_\eta$ and critical current $I_s$ assuming $100\%$ spin polarization because the polarization factor is essential in determining the asymmetry in P-to-AP and AP-to-P switching and it is from the material properties independent of the macrospin error model. In the simplest case, one can just relate those quantities through $I_\eta=I\eta$ and $I_s=I_c\eta$ where $\eta$ is the polarization of the fixed magnet. Once the WER and $t$ are given, the equation solves for $i$ that matches the design target (an example of such a plot, obtained from numerical simulation of Fokker-Planck, was shown earlier in Fig.\ref{fig:WER-t}). The second step is to translate the scaled $i$ into the polarized charge current $I_\eta$. To do so, we need the intrinsic critical current $I_s$ assuming $\eta=100\%$ spin polarization, which we can solve under the macrospin assumption\cite{Sun2000}. Notice that these two modules have taken into account most of the magnetic properties of the free magnetic layer including the saturation magnetization, magneto-anisotropy and magnetic damping . To further evaluate the energy dissipation, a transport model is required to relate the polarized charge current to the voltage across the MTJ. While this needs a full NEGF simulation to capture the spin dependent electron tunneling, we use instead a simple transport model based on the modified Simmons equation derived in \cite{munira2012quasi}. In that free electron model, the electronic structure of an MTJ can be described by the following six parameters: 1) the effective mass of the ferromagnetic layers $m_c$; 2) the effective mass of the nonmagnetic spacer $m_{barr}$; 3) the Fermi energy to the bottom of the conduction band $E_F$; 4) the potential barrier height $U$; 5) the thickness of the nonmagnetic layer $d$; and 6) the band-splitting in the ferromagnetic layers due to the exchange energy $\delta$. Under the free electron assumption, $k^\mathrm{p}_{\uparrow\downarrow}=\sqrt{2m_c[E_F-(\delta\mp\delta-qV)/2]}/\hbar$ and $k^\mathrm{f}_{\uparrow\downarrow}=\sqrt{2m_c[E_F-(\delta\mp\delta+qV)/2]}/\hbar$ are the spin dependent wavevectors in the pinned layer and the free layer respectively, $\beta$ is the tunneling parameter $\beta=m_\mathrm{c}\kappa/m_\mathrm{barr}$ and $\theta$ is the angle between the magnetic moments of the pinned layer and the free layer. These six material parameters can be adjusted to fit the experimental TMR data at low-bias. Once fitted, they can simulate the entire I-V characteristics and spin transfer torque as well \cite{munira2012quasi}. It also can be used to predict additional properties such as the spin torque switching voltages for example. The Simmons model incorporates the bandstructure from which the polarization factor can be extracted (Fig.\ref{fig:diagram} Block 3).  From the I-V (Fig.~\ref{fig:diagram} Block 4) and $\eta(V)$, one can plot the $I_\eta(V)$, which combined with the input $I_\eta$, can determine the switching point and switching voltage as shown in Fig.~\ref{fig:diagram} (Block 2) where two `jump points' indicate the switching between parallel and anti-parallel states. The final output from this process is the energy dissipation of STT switching in MTJs. 

Even though each equation shown in diagram Fig.\ref{fig:diagram}
has its limitations, the idea of such integrated picture is more general and important. For MTJs, each block can be easily replaced by a more accurate and appropriate numerical model if needed. Some can even come from experimental measurements, which usually only focus on of those blocks. Putting those data in the context of other properties or requirements gives a clearer picture of where the technology stands. For other nanomagnetic applications, one can formulate the corresponding equations/models and build a similar process.

\section{Towards System Level Simulations}
\label{sec:circuit_modular}

\subsection{Modular Approach to Spintronics}
In the previous sections of this paper, we have demonstrated how the physics of charge and spin transport combined with magnetization dynamics of nanomagnetic materials can be modeled starting from ab-initio methods such as DFT. However, a multi-scale approach that can cater to an increasing number of materials and phenomena relevant for spintronics calls for the ability to create abstractions for `lower level' physics based models and reuse those models and techniques for more complex designs built with multiple materials and phenomena, without losing the accurate behavior and properties captured by those models. 

A recently developed Modular Approach to Spintronics \cite{camsari2015modular}, based on the multi-component spin-circuit formalism, provides such a framework for multi-physics, multi-scale modeling, and simulation of spintronic and nanomagnetic devices. In essence, the approach is based on a set of carefully benchmarked elemental "circuit modules" for transport physics and magnetodynamics and interaction through various materials. These models are then combined in a `lego-block' fashion to build a larger circuit model for a complex experimental structure or functional device and then simulated in a standard circuit simulation software (SPICE). This approach was used to study the performance and dissipation in a large family of spintronic devices as well as explore novel non-Boolean computing schemes using stochastic magnetodynamics \cite{ganguly2017evaluating}.

In a nutshell, the aim of this effort is to marry decades of development in complex circuit simulation techniques with fundamental atomistic methods surveyed in this paper. The Modular Approach to Spintronics is a multi-organization open source initiative with a large library of modules and illustrative example models that are available through its project portal \cite{modularwebsite}.

A typical workflow utilizing the methods laid out in this paper will be:
\begin{enumerate}
\item Materials Modeling (DFT): calculate stable candidate materials and their material properties such as anisotropy. Generate Hamiltonian matrices for the material for use in transport calculations.
\item Transport Modeling (NEGF, Spin-Diffusion): Using the Hamiltonian generated from materials modeling, apply NEGF methods to calculate I-V and other transport characteristics (say torques, spin currents etc.)
\item Compact Modeling: From the transport modeling results, develop analytical/semi-analytical equations relating charge and spin currents and voltages at the `terminals' of the material.
\item Device Modeling: Use the compact model developed before to create a circuit model for the test device, bringing together other modules previously developed, which can also include magnetodynamic modules (LLG).
\item Circuit Modeling: Use the device model in a circuit testbench of choice and analyze performance. Use the results to define the targeted parameter space and go back to step 1 to search for `better' materials.
\end{enumerate}

\subsection{Spin-based Logic}
The emerging write mechanisms discussed above have created the possibility of building multi-terminal nanomagnetic devices where simultaneous read and write operations can be performed, unlike a two-terminal MTJ. As a result, a wide variety of spintronic logic devices have been proposed. The fundamental principle underlying most of these devices is the coupling of a write and a read unit \cite{datta_what_2014}, which may or may not be electrically isolated. Some of the spin-logic device proposals are:

\begin{itemize}
\item All Spin Logic (ASL) \cite{behin-aein_proposal_2010} and its variants: Works on the principle of non-local spin-transfer torque, where two nano-magnets interact through non-local spin-currents. The variant designs attempt to improve the performance by using different materials for the spin-channels, PMA magnets, VCMA effect to assist in writing, etc.
\item Domain Wall based spin-logic devices (including m-Logic \cite{zhu_mlogic:_2015}): Works on the principle of the controlled motion of a magnetic nanowire's domain through spin-torque and read-out by an MTJ. It can be viewed as an ASL device using spin-waves instead of spin-currents.
\item Spin-Switch using Spin-Orbit torques \cite{datta_non-volatile_2012}: Using a GSHE driven magnet as a write unit and a complementary MTJ pair as a read unit, with both coupled by a magnetic interaction. In the rest of this section, we will show how to build circuit model for the Spin-Switch using the Modular Approach.
\item Magnetoelectric spin logic devices \cite{ganguly2017evaluating,jaiswal2017mesl,mankalale2016fast}: Works by using the magnetoelectric effect as the writer for a spin logic device.
\item Strain-based devices \cite{dsouza_four-state_2011}: Works by using the piezzoelectric effect to manipulate the anisotropy of the nanomagnet and consequently reducing the write current in a spin-logic switch.
\item Nano-Magnetic Logic \cite{imre_majority_2006}: QCA like computing with nano-magnets interacting through magnetic fields rather than spin-currents.
\end{itemize}

\subsection{Circuit Model for the Spin-Switch and Device Characteristics}
The Spin-Switch \cite{datta_non-volatile_2012} is composed of a write unit, consisting of a GSHE-driven free nano-magnet, and a read unit, consisting of a complementary MTJ pair (AP$\mid$P) sharing a common free layer. The two free layers are coupled magnetically with an oxide layer that provides electrical isolation between the two units. This device has an internal gain from the GSHE based writer, which ensures that it is switched by the write magnet in preference to the read magnet since the MTJ pair's spin-current does not have any gain, providing directivity of information flow from input to output. The electrical isolation allows the separation of biasing points among various stages in a circuit built from this device.

While the spin switch has all the required features for a logic device, its performance is not comparable to a CMOS inverter in terms of energy-delay product per switching event metrics \cite{nikonov_benchmarking_2015} due to inefficiencies associated with spin-torque switching. But it can be improved to a large extent by adopting various high-performance materials \cite{ganguly2017evaluating}. Here we do not go into the details of the physics of the spin switch and various optimization possibilities. Instead, we want to show how to assemble a device model for the Spin-Switch (available from \cite{nanoHUB.org21}) and explore its operations.

\begin{figure}[ht]
\includegraphics[width=8.5cm]{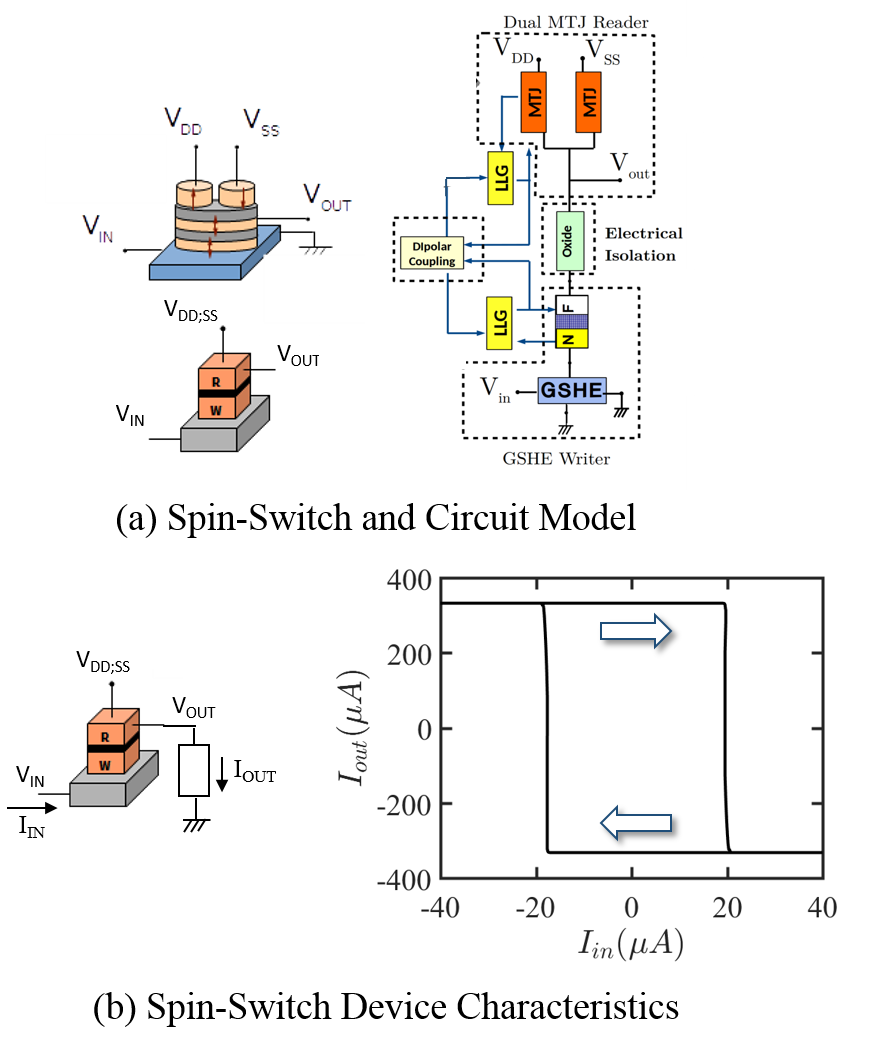}
\caption{ (a) Schematic of the Spin-Switch device and its circuit model built using the Modular Approach to Spintronics (b) Circuit testbench to measure the device characteristics. Once setup, various different characteristics can be obtained. For example, the Input current vs. the Output current is shown here.}
\label{fig:spin_switch}
\end{figure}

Fig. \ref{fig:spin_switch}a shows the schematic design of the  Spin-Switch on the top left and its compact representation at the bottom left. On the right, we show the circuit models based on the elemental modules for the transport and the magnetodynamics using the Modular Approach, e.g. the GSHE module encompasses the physics of the charge/spin transport through the GSHE material in the Spin-Switch design. The LLG module calculates the instantaneous time dependent magnetization of a nanomagnet with two inputs: the magnetic field and the injected spin current. The magnetization can then be fed into the magnet module and the MTJ module, as well as the magnetic interaction module (dipolar in this case), to calculate the total mutual magnetic field between the read and the write units. 

The modules have been designed to maintain the validity of Kirchoff's Laws of circuits by treating phenomena such as spin generation and decay as dependent sources and shunts, and the non-electrical quantities such as spin current, voltages, magnetic fields, and magnetizations as vector voltages and currents. The circuit model can, therefore, be connected with any other classic electrical elements like voltage and current sources, capacitors, resistors, transistors etc. and used in any SPICE-like circuit simulator to perform sophisticated simulations such as parametric and data-driven sweeps, noise, measure based metrology, rf and etc.

As an example, we use a simple testbench shown in fig. \ref{fig:spin_switch}b. The input current is swept first from positive to negative values and then vice versa. The output current flowing through the load resistor is recorded. The device characteristics show hysteresis, as expected from a device with built-in memory from the magnets. It is clear that the swing of the output current is larger than the input current window, which indicates a gain in the device that enables it to drive multiple Spin-Switches at its output, i.e. fan out $>$ 1, and can be used to build logic pipelines without the need of amplifier or buffer stages. The output load was chosen to match the input impedance of another Spin-Switch (an arbitrary choice) and the Spin-Switch's magnetic anisotropy barrier is set $\Delta = 40\ kT$, which is not necessary for a logic device but has been chosen to demonstrate the built-in memory aspect of the device. The sweeps are performed in a time varying fashion, i.e. a transient analysis has been used instead of the dc sweep to accurately capture the magnetodynamics of the device. Thermal noise has been ignored for this simulation, i.e. the simulation is performed at $T=0K$.

\subsection{Simulating Noise Margin and Signal Recovery in a Logic Pipeline}

The Spin-Switch circuit model can be used to build larger circuits and analyze their performance through the circuit simulation. See \cite{gangulythesis} for a comprehensive study of Energy-Delay of the Spin-Switch and its variant designs. In this section, we will use a FO-1 chain of inverters built from the Spin-Switch to study its noise margin and recovery. 

\begin{figure}[ht]
\includegraphics[width=8.5cm]{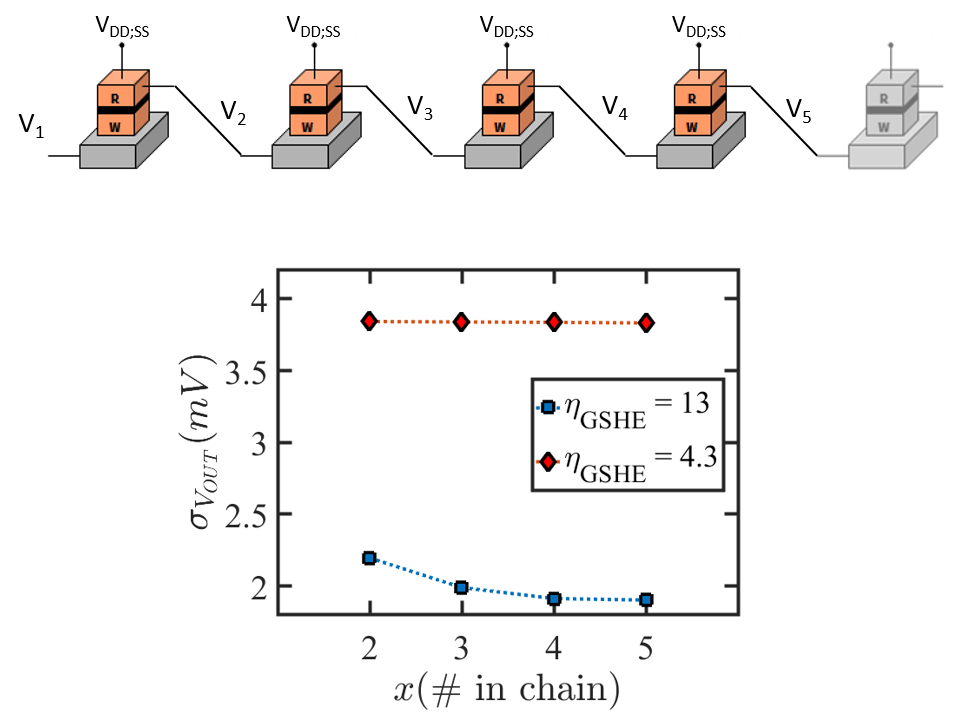}
\caption{Simulating the error propagation and signal recovery in a FO-1 Spin-Switch based inverter chain for two different Spin-Switch devices.}
\label{fig:spin_switch_ckt}
\end{figure}

A chain of 4 inverters built from identical Spin-Switches with a terminating load matching the input impedance has been setup as shown in fig. \ref{fig:spin_switch_ckt}. Using transient noise simulation available in commercial SPICE simulators, it is possible to simulate the Langevin dynamics of the nanomagnets (Brownian motion). This is done by including a noise source in the LLG solver module and record the noisy output from the devices, consistent with how an experiment may be performed in a real life setup. It is then possible for a Monte Carlo simulation of the dynamics over many samples to obtain the statistical properties matching the real device performances.

We perform a transient simulation for slowly rising input signal $V_1$ and record the switching points of each Spin-Switch in the chain ($V_{2,3,4,5}$). From the multiple Monte Carlo samples, we calculate the standard deviation of the switching voltages for each device. This calculation is performed with two different GSHE gains.

It can be seen that for the device with a higher gain, the overall deviation is lower compared to the device with a lower gain, resulting in a tighter noise margin. This can be explained by the fact that higher gain helps boost the Signal-to-Noise ratio (noise being thermal in origin). Additionally, it can be seen that the deviation decreases further down the chain. This shows that the gain helps in recovery of a noisy input signal as it propagates through a logic pipeline, and higher the gain, better the recovery. 

The point of this exercise is, firstly, to show a proof-of-concept complex simulation made available for spintronic devices due to the Modular Approach. Secondly, it demonstrates the principle of signal recovery through the in-built gain in a circuit composed of multiple Spin-Switches.

\subsection{Probabilistic Spintronic Logic}

A recent development in spintronic logic is to utilize the stochastic switching of nanomagnets for novel non-Boo\-lean circuits and architectures. As noted earlier, the spin-torque based switching is inherently disadvantageous compared to MOSFETs in terms of Energy-Delay-Reliability product. Therefore it does not seem to be a viable alternative as a direct CMOS drop-in replacement in conventional circuits.

However, there are other possible applications where the inherent physics of spintronics and nano-magnets can enable new classes of devices with higher functional efficiency, i.e. to map a complex logic function directly into the hardware. There are two attractive features of spintronic/nanomagnetic devices: the natural addition of currents in metal interconnects of spin devices that enables the majority logic type functionality; the thresholding of magnetization switching that is attractive in building neuromorphic circuits \cite{diep_spin_2014,sharad_spin-neurons:_2013,ramasubramanian_spindle:_2014}. In addition, stochastic thermal switching of nanomagnets has opened the possibility of building networks that embrace the uncertainties for more energy efficient circuits \cite{ganguly2017evaluating,us_department_of_commerce_stochastic_????,venkatesan_spintastic:_2015,locatelli_spintronic_2015,behin2016building,PSLpaper,faria2016low}.

\begin{figure}[ht]
\includegraphics[width=8.5cm]{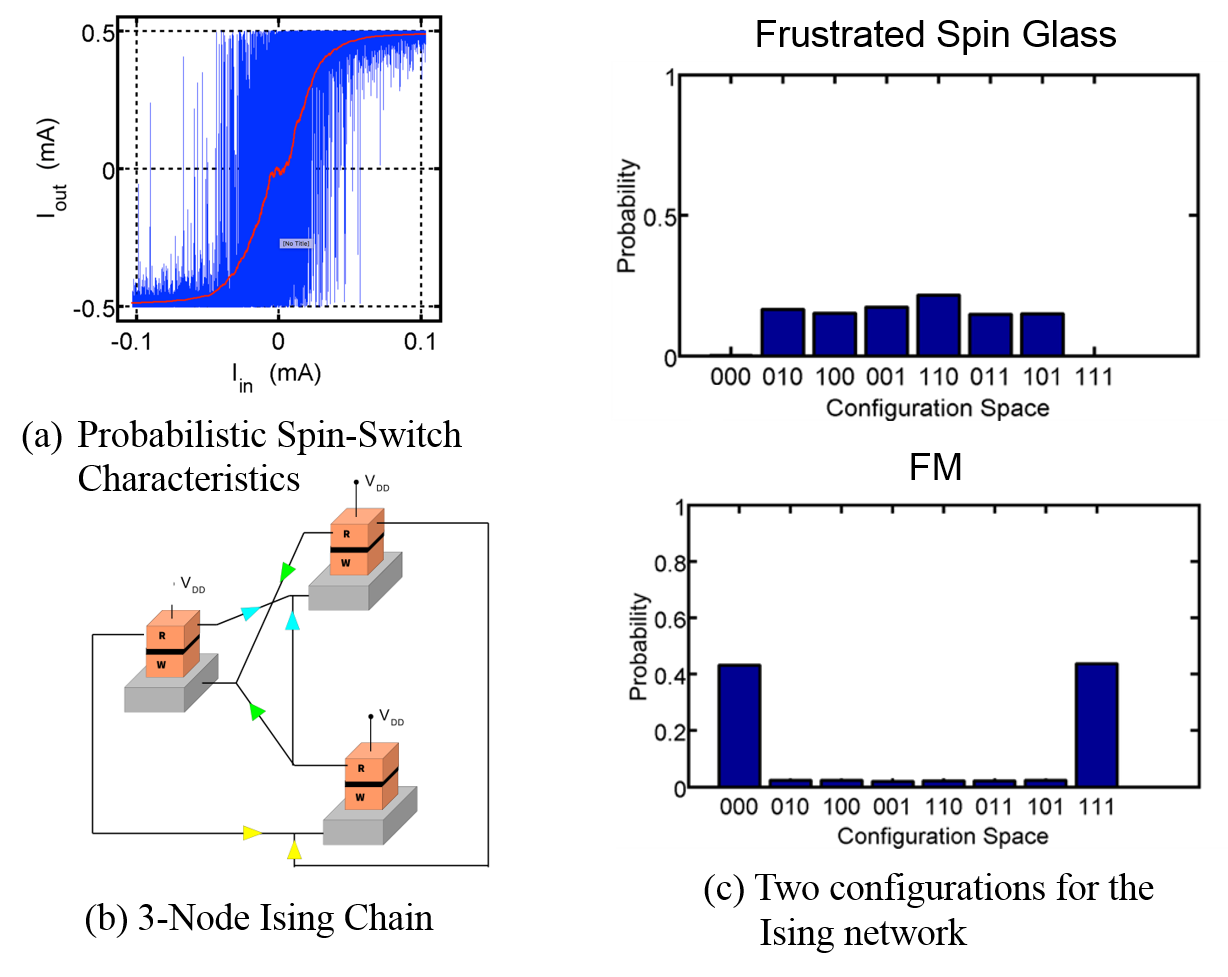}
\caption{ (a) Device characteristic of a stochastic Spin-Switch made of superparamagnets instead of ferromagnets. The instantaneous response (blue) of the device is "noisy" but its statistical properties (mean, red curve in the foreground) can be controlled by the input current. (b) A 3-node Recurrent Boltzmann Machine built using the Stochastic Spin-Switch. The interactions are controllable and are programmed to mimic an Ising chain. (c) The two configurations of the Ising chain, Ferromagnetic (FM) and Frustrated Spin Glass. Adapted from \cite{ganguly2017evaluating}. \textcopyright IEEE 2016}
\label{fig:stochastic}
\end{figure}

In the rest of this section, we illustrate the principles behind the spintronic stochastic switching and computation using a variant of the Spin-Switch design. For this example, we lower the anisotropy barrier of the nanomagnet in the Spin-Switch from $\Delta = 40\ kT$ to $\Delta \sim 2\ kT$, which can be achieved with a superparamagnet. From eq. \ref{eq:tau_magnet}, the lower energy barrier reduces the device retention from a decade to a few ns. It can be shown \cite{ganguly2017evaluating,sutton2016intrinsic} that the statistical mean of the output voltage can be controlled by a spin-current, which produces a device characteristic that resembles a sigmoid function(see fig.\ref{fig:stochastic}a. The blue background is the noisy instantaneous response as the input current is swept, while the red is a sampled mean produced through a $R-C$ network acting as a low pass filter.) 

Connecting these devices together with controllable interaction through spin current allows us to form Restricted Boltzmann Machines (a class of stochastic recurrent neural networks) that can learn and reproduce patterns through annealing. The stochasticity of the superparamagnets is advantageous in scanning the phase space at GHz frequencies and reaching the ground state (solution state) quickly during the annealing process.

A proof-of-concept example of the annealing process is shown in fig. \ref{fig:stochastic}b, where a circuit built from 3 stochastic Spin-Switches forms a 3-node Ising network. The connections among them can be weighted by an external circuitry independent from the Spin-Switches. Some possible ways to implement the weighted connections are voltage controlled resistors such as memristors, bias voltages on the Spin-Switches themselves, or an external CMOS based circuitry \cite{borders2016analogue}. In this example, the magnitudes of the weights are chosen to give strong connections that can drive the devices unambiguously into the saturation regions in fig.\ref{fig:stochastic}a. When the signs of the connections are all positive, the interaction type is ferromagnetic and therefore after annealing, those devices prefer $000$ and $111$ states equally, with very low probability to choose other states. On the other hand, if the interactions are negative (anti-ferromagnetic) the networks settle into a frustrated spin-glass state (fig.\ref{fig:stochastic}c).

With this type of network, it is possible to compute any logic function by adjusting the interactions to encode the desired answer or truth table into its eigenstates, which the network can find quickly through the annealing process. An example is to solve a Travelling Salesman Problem (a classic NP-complete problem) with an Ising network demonstrated in ref. \cite{sutton2016intrinsic}. Another example is to implement a 32-bit Ripple Carry Adder with functional complementarity discussed in ref. \cite{PSLpaper}.

The purpose of this section is to demonstrate the capabilities of the integrated DFT to SPICE approach presented in this paper that allows us to directly connect the physics of the materials to the performance of functional memory and logic devices as well as explore novel circuit applications enabled by these materials.

\section{Summary}
Traditional STT-MRAMs have already started commercializing to some extent. The continual improvement in the material engineering and fabrication process will promise better device performance in terms of the read-write-reliability metrics, which makes it a competitive candidate in memory applications. Following these efforts, we have summarized the computational tools that facilitate the material study and demonstrated a wide range of material choices in the Heusler family with the potential to improve the performance of MTJs at scaled nodes. We have also illustrated the importance of the thermal noise in magnetization switching and how to characterize its effect using the LLGS and Fokker-Planck methods. A proper engineering of the nanomagnet (such as tilted magnetic moment) can mitigate some negative effects from the thermal noise. However, the low power dissipation goal promised by nano-magnetics can hardly be achieved in conventional magnetic tunnel junction structure with spin transfer torque switching. On the one hand, emerging device ideas such as GSHE bring new hope toward energy efficient operations of nanomagnetic devices in traditional applications. On the other hand, a paradigm shift towards non-boolean computation might bring nanomagnetic into new application realms such as probabilistic computing, neuromorphic computing and etc. In any case, a comprehensive understanding of a nanomagnetic device - from its material to its circuit performance is necessary to be able to optimize it to meet the requirements of different applications. To do that, we have shown an integrated framework that connects the material properties to the circuit performance and illustrated it with different devices such as the STT-MRAM and the Spin-Switch. Each part of the toolbox is general enough to be extended to study other device ideas. We hope these tools can serve researchers from different communities and connect their expertise and advancement to inspire novel yet practical solutions.


%



\section*{Acknowledgment}
We acknowledge discussions with Prof. W. H. Butler on Heusler studies, I. Rungger and S. Sanvito from the {\it Smeagol} team for ab-initio transport calculations, P. Visscher and B. Behin-Aein on Fokker-Planck and write error study in STT-MRAM. We also acknowledge the support from Oak Ridge National Laboratory (ORNL) center for nanophase materials sciences (CNMS) on computational resources. This work was funded over the years by NSF-DMREF-1235230, DARPA, NSF-SHF, SRC-NRI and IBM Fellowship. 

\bibliographystyle{unsrt}

\end{document}